\documentclass[3p,times]{elsarticle}

\usepackage{ecrc}

\volume{00}

\firstpage{1}

\journalname{Icarus}

\runauth{}

\jid{Icar}

\jnltitlelogo{Icarus}

\usepackage{amssymb}

\usepackage[figuresright]{rotating}

\begin{document}

\begin{frontmatter}



\dochead{}

\title{How to link the relative abundances of gas species in coma of comets to their initial chemical composition ?}


\author[label1,label2]{Ulysse~Marboeuf}
\author[label2]{Bernard~Schmitt}
\address[label1]{Physics Institute and Center for Space and Habitability, University of Bern, Bern, Switzerland}
\address[label2]{UJF-Grenoble 1, CNRS-INSU, Institut de Plan{\'e}tologie et d'Astrophysique de Grenoble (IPAG), UMR 5274, 38041 Grenoble, France}

\begin{abstract}
Comets are expected to be the most primitive objects in the solar system. The chemical composition of these objects is frequently assumed to be directly provided by the observations of the abundances of volatile molecules in the coma.
The present work aims to determine the relationship between the chemical composition of the coma, the outgassing profile of volatile molecules and the internal chemical composition, and water ice structure of the nucleus, and physical assumptions on comets.
To do this, we have developed a quasi 3D model of a cometary nucleus which takes into account all phase changes and water ice structures (amorphous, crystalline, clathrate, and a mixture of them);
we have applied this model to the comet 67P/Churyumov-Gerasimenko, the target of the Rosetta mission.
We find that the outgassing profile of volatile molecules is a strong indicator of the physical and thermal properties (water ice structure, thermal inertia, abundances, distribution, physical differentiation) of the solid nucleus. Day/night variations of the rate of production of species helps to distinguish the clathrate structure from other water ice structures in nuclei, implying different thermodynamic conditions of cometary ice formation in the protoplanetary disc. The relative abundance (to H$_2$O) of volatile molecules released from the nucleus interior varies by some orders of magnitude as a function of the distance to the sun, the volatility of species, their abundance and distribution between the "trapped" and "condensed" states, the structure of water ice, and the thermal inertia and other physical assumptions (dust mantle, ...) on the nucleus.
For the less volatile molecules such as CO$_2$ and H$_2$S, the relative (to H$_2$O) abundance of species in coma remain similar to the primitive composition of the nucleus (relative deviation less than 25\%) only around the perihelion passage (in the range -3-2 to +2-3 AU), whatever is the water ice structure and chemical composition, and under the conditions that the nucleus is not fully covered by a dust mantle. 
The relative (to H$_2$O) abundance of highly volatile molecules such as CO and CH$_4$ in the coma remain approximately equal to the primitive nucleus composition only for nuclei made of clathrates. The nucleus releases systematically lower relative abundances of highly volatile species (up to one order of magnitude) around perihelion (in the range -3-2 to +2-3 AU) in the cases of the crystalline and amorphous water ice structures in the nuclei.
The rate of production, the outgassing profile and the relative abundances (to H$_2$O) of volatile molecules are the key parameters allowing one to retrieve the chemical composition and thermodynamic conditions of cometary ice formation in the early solar system. The coming observations of the coma and nucleus by the Rosetta mission instruments (VIRTIS, MIRO, ...) should provide the necessary constraints to the model to allow it to infer the primordial ice structure and composition of the comet.
\end{abstract}


\end{frontmatter}



\section{Introduction}

Comets are expected to be the most primitive bodies in the solar system. It is believed that these objects are the witnesses of the formation of the solar system, and that their study could help to understand the conditions of formation and evolution of the primitive solar system. The study of these objects is crucial to determine the chemical composition and the thermodynamic conditions of ice formation in the protoplanetary disc and the early (primitive) solar system. Observations of these bodies (Bockel{\'e}e-Morvan et al.~2004; Mumma \& Charnley 2011) show variations of abundances of all the species (relative to H$_2$O) up to 2 orders of magnitude (see A'Hearn et al. 2012) whatever the position of comets around the sun (Note that 'abundance' in this paper refers to relative (to H$_2$O) production rates rather than mixing ratios). 
These variations may be due either to the dynamic and collisional history of comets, or to a different initial chemical composition of these objects linked to a different area of formation in the nebula disc, before or after the respective ice lines of the volatile species. 
For each comet, these variations are also a function of the distance to the sun as shown for comet Hale-Bopp (Biver et al. 2002). This observation could be explained to first order as follow: the volatile species sublimate at different rates as a function of the temperature, i.e. the distance to the sun. However Marboeuf et al. (2012) have shown from a 1D cometary model that the rates of production of the volatile species and their outgassing profiles\footnote{The outgassing profile refers to an evolution of the outgassing with the heliocentric distance} are mainly function of the thermal inertia of the cometary materials, the nature of the water ice structure (amorphous, crystalline, or clathrate type structure), and the abundance and distribution of volatile molecules between the `trapped' and `condensed' states in comets.
The present work aims to determine the relationship between the abundance of gas species in the comae of comets (the main observational information on these objects) and the primitive internal abundance of ice species within the nucleus. In particular, we will study the effects of the physical and thermodynamical properties such as the water ice structure (amorphous, crystalline, clathrate and a mixture of the three), the thermal inertia, the abundance and distribution of species between the `trapped' and `condensed' states, and the presence of a dust mantle on the surface of nuclei on the relative abundances and the outgassing profiles of volatile molecules at the surface of comets. The main objective of this study is to constrain some general observational keys for the interpretation of outgassing observations of comets, in particular the future one of the comet 67P/Churyumov-Gerasimenko, the target comet of the Rosetta mission, and the one of the comet Hale-Bopp.

This article is organized as follow: Sec.~\ref{sec:models} is devoted to the presentation of the quasi 3D model of a cometary nucleus and the description
of the main physical processes taken into account. In Sec.~\ref{paramsection}, we discuss the physical assumptions, chemical composition and thermodynamics parameters adopted for the nucleus. In Sec.~\ref{results} we present results about the physical differentiation, rate of production, outgassing profile and relative abundances of the species as a function of the physical properties (thermal inertia, water ice structure, chemical composition, dust mantle ...) of the cometary nucleus. Sec.~\ref{observations} is devoted to the comparison of models with some observational data.
We finally discuss and summarize the main results in Sec.~\ref{discussion}.

\section{Quasi-3D model of comet \label{sec:models}}

\begin{figure}
\includegraphics[width=10.cm]{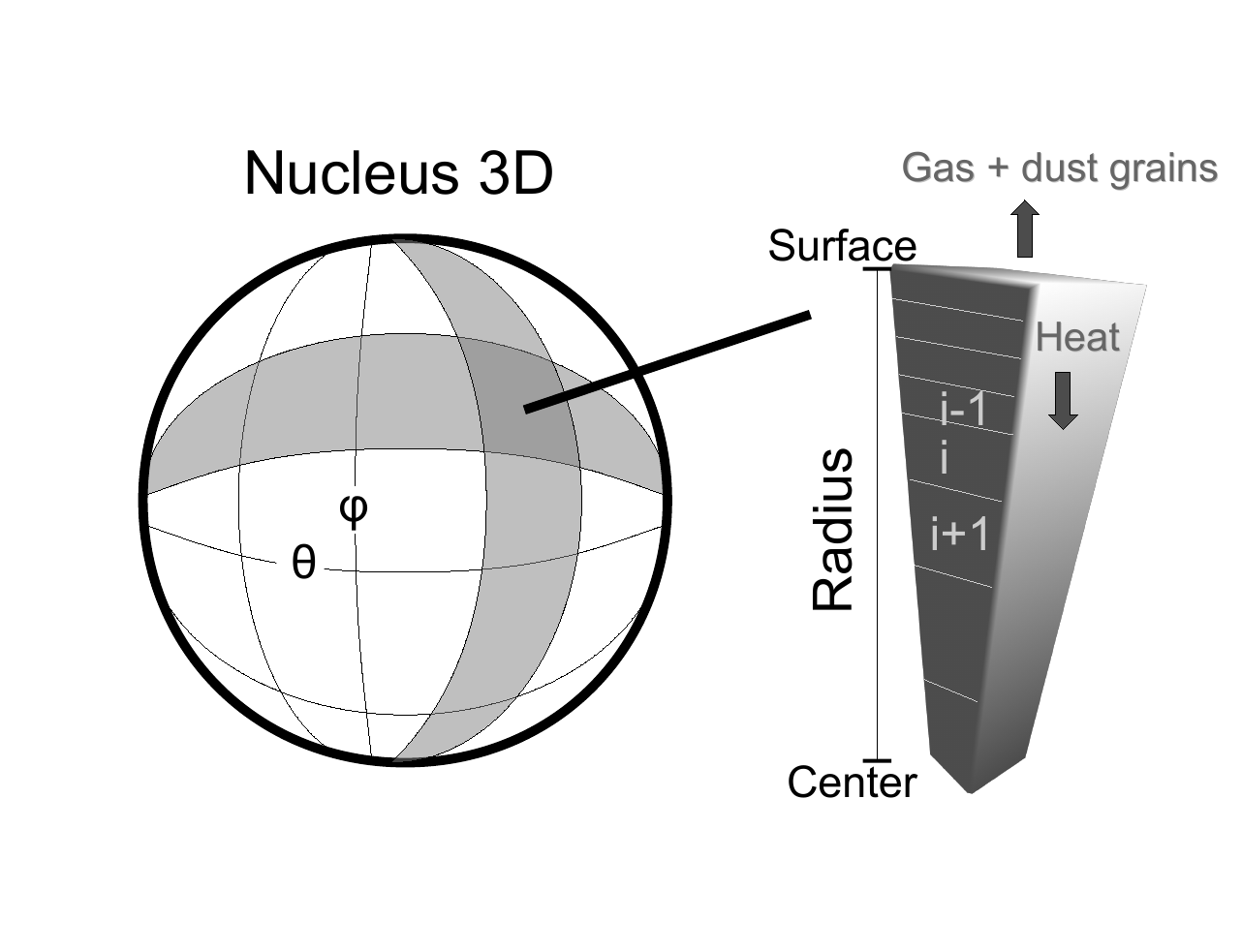}
\caption{Schematic view of the quasi 3D nucleus model of comet. Heat conduction and gas diffusion occur only radially throughout the nucleus.}
\label{fig:3D}
\end{figure}

The model simulates the cometary material as an icy porous matrix composed of dust grains with an icy mantle formed of water and some other volatile species in solid states (see Fig.1 in Marboeuf et al. 2012). 
The numerical model presented in this work uses the quasi 3D approach. 
The quasi 3D approach allows us to take into account spatial (latitudinal and longitudinal) variations of the temperature on the surface of nuclei due to the unevenly distributed solar radiation over the cometary surface (see Weissman \& Kieffer 1981; Cohen et al. 2003; Lasue et al. 2008). This leads to a much better estimation of the temperature on its surface compared to the 1D model (Marboeuf et al. 2012) which considers an average temperature everywhere on the surface of the comet with spherical symmetry whatever the erosion of the nucleus.
This model represents a spherical nucleus whose surface is divided numerically in several sections ($\varphi$, $\theta$) as illustrated in
Fig.~\ref{fig:3D}, and below which the interior of the nucleus is divided in several radial layers ($i$ index) whose thickness follows initially a power law (see right part of the figure ~\ref{fig:3D}). The size and number of these radial layers can increase or decrease during the lifetime of the comet following its erosion (sublimation of ice and dust grain ejection). Note that each section ($\varphi$, $\theta$) evolves independently of the others and only radial flow of gas and heat are considered in the model.

The model takes into account several volatile species together (H$_2$O, CO, CO$_2$, ...) and several types of water ice structures: amorphous with trapped gases, pure crystalline, clathrate with trapped gases or a mixture of these structures, depending on the formation location of the comet in the solar system and assumptions on the origin of the cometary material (see Marboeuf et al., 2012).
Within the cometary nucleus, the model describes radial heat transfers, latent heat exchanges, H$_2$O ice phase transitions (amorphous $\rightarrow$ crystalline, crystalline $\leftrightarrow$ clathrate, and amorphous $\rightarrow$ clathrate), sublimation/condensation of volatile molecules in the porous network of the nucleus, radial gas diffusion, as well as the allowed gas releases/trapping by/in the water ice structures. 

At the surface of the nucleus, the model takes into account the gases and dust grains ejections as well as a dust mantle formation. Descriptions and assumptions on the physical processes taken into account in the model are fully explained in Marboeuf et al. (2012).
Hereafter, we provide a simple description of the main physical processes taken into account in the model. In order to ensure perfect conservation of mass and energy in the model, we use finite volume method (Patankar 1980) for the discretization of equations (\ref{NRJ_cons}) and (\ref{MASS_cons}) explained hereafter.

\subsection{Energy conservation}

For each layer $i$ and position ($\varphi$, $\theta$) in the nucleus, we solve the energy conservation equation that describes the radial heat diffusion through the porous matrix:
\begin{equation} 
  \rho c \frac{\partial T}{\partial t}= \nabla. \left(K^{m} \frac{\partial
 T}{\partial r}\right) - \sum_{x} H_x^s Q_{x} +Y   \qquad (J \, m^{-3} \, s^{-1}) 
\label{NRJ_cons}
\end{equation}
with $\rho$ (kg m$^{-3}$) the density of solids, $c$ (J kg$^{-1}$ K$^{-1}$) their specific heat capacity, $T$ (K) the temperature, $t$ (s) the time, and $r$ (m) the distance to the center of the nucleus. $H_x^s$ (J mol$^{-1}$) is the molar latent heat of sublimation of ice $x$ and $Q_x$ (mol m$^{-3}$ s$^{-1}$) represents the rate of moles of gas $x$ per unit volume that sublimates$/$condenses in the porous network or/and that is released by amorphous ice during the process of crystallization. Its expression is given by the inversion of the gas diffusion equation~(\ref{MASS_cons}) given below.
$Y$ represents the power per unit volume released during the crystallization process of amorphous water ice (see Espinasse et al. 1991), exchanged between the gas phase (which diffuse in the porous network) and the solid matrix, or/and released/taken during the formation/dissociation of cages of clathrate.
$K^{m}$ (J s$^{-1}$ m$^{-1}$ K$^{-1}$) is the heat conduction coefficient of the porous matrix whose expression is given by Hertz's formula (see Kossacki et al. 1999; Davidsson \& Skorov 2002; Prialnik et al. 2004; Huebner et al. 2006; Marboeuf et al. 2012):
\begin{equation}
K^{m}=h K^s  \qquad \textstyle {(W \, m^{-1} K^{-1})}
\label{ctc}      
\end{equation}
where $K^s $ is the conductivity of the solid (dust and ices) components (see Marboeuf et al. 2012). $h$ is the Hertz factor used to correct the effective area of the matrix material through which heat flows (Davidsson \& Skorov 2002; Prialnik et al. 2004). $h$ is expressed by considering two spheres of radius $R$ that are pressed together and have a contact area of radius $r_c$ ($h\approx \frac{r^2_c}{R^2}$, Kossacki et al. 1999). Its value can vary between 10$^{-3}$ and 10$^{-1}$ (see Davidsson \& Skorov 2002; Huebner et al. 2006, Volkov \& Lukyanov 2008).

\subsection{Mass conservation}
For each layer $i$, position ($\varphi$, $\theta$) and molecule $x$, we solve the diffusion of gas through the porous matrix by using the mass conservation equation:
\begin{equation}
  \frac{\partial \rho_x^g }{\partial t}=M_x \left(\nabla .\left (\Phi_x \right) +Q_x \right) \qquad (kg \, m^{-3} \, s^{-1})
\label{MASS_cons}
\end{equation}
where $Q_x$ (mol m$^{-3}$ s$^{-1}$) represents the net source of gas $x$ released in the porous network during water ice crystallization, the pure ice sublimation/condensation, and/or the rate of clathrate dissociation/formation. $\rho_x^g$ (kg m$^{-3}$) is the mass density of gas $x$, and $M_x$ (kg mol$^{-1}$) its molar mass. $\Phi_x$ (mol m$^{-2}$ s$^{-1}$) is the molar flow through the porous network whose expression is given in Marboeuf et al. (2012).

\subsection{Calculation of the temperature and sublimation of ices at the surface of the comet}

The sublimation of volatile molecules at the surface of the nucleus is mainly a function of the temperature $T_{\varphi, \theta}$ of the section of surface at the position ($\varphi$, $\theta$). This temperature is given by a thermal balance between the solar energy absorbed by the cometary material (on the left side of the equation) and its thermal emission, the heat diffusion towards the interior of the nucleus and the energy of sublimation of ice species (on the right side of the equation) on the elemental surface ($\varphi$, $\theta$):
\begin{equation} 
\frac{C_s(1-A_b)}{r_h^2} ~\zeta_{\varphi, \theta} = \epsilon \sigma T_{\varphi, \theta}^4 + K \frac{\partial T_{\varphi, \theta}}{\partial r} + \sum_x \alpha^i_x H_x^s \varphi_x(T_{\varphi, \theta}) \qquad (W \, m^{-2})
\label{boundary_heat_surf}
\end{equation}
where $C_s$ (W m$^{-2}$) is the solar constant, $A_b$ the Bolometric Bond Albedo, $r_h$ (AU) the distance to the sun, $\epsilon$ the infrared surface emissivity, $\sigma$ the Stefan-Boltzmann constant (W m$^{-2}$ K$^{-4}$), $T_{\varphi, \theta}$ (K) the temperature of the surface, $\alpha^i_x$ the surface fraction covered by the pure ice specie $x$ (including H$_2$O). $\varphi_x$ (mol m$^{-2}$ s$^{-1}$) is the free sublimation rate of specie $x$ in vacuum given by the expression of Delsemme $\&$ Miller (1971):
\begin{equation}
  \varphi_x(T_{\varphi, \theta})=\frac{P_x^s(T_{\varphi, \theta})}{\sqrt{2\pi M_x~R~T_{\varphi, \theta}}} \qquad (mol \, m^{-2} \, s^{-1})
\end{equation}
where $M_x$ (kg mol$^{-1}$) is the molar mass of the corresponding gas specie, $P_x^s(T_{\varphi, \theta})$ its vapor sublimation pressure (Pa), and $R$ the
perfect gas constant (J mol$^{-1}$ K$^{-1}$).\\
$\zeta_{\varphi, \theta}$ is the term of insolation at the position ($\varphi$, $\theta$) on the surface of the nucleus.
We use the "slow-rotator" approach that takes into account the diurnal latitudinal and longitudinal variations of illumination on the facets of the surface. This approach allows one to obtain an accurate surface temperature distribution and its diurnal changes at any heliocentric distance (Prialnik 2004).  With this approach, $\zeta_{\varphi, \theta}$ is equal at $max(\cos~\xi, 0)$ in the Eq.~(\ref{boundary_heat_surf}), with $\xi$ the stellar zenith distance calculated as (see Sekanina 1979; Fanale \& Salvail 1984; Prialnik 2004; Gortsas et al. 2011; Marboeuf et al. 2012):
\begin{equation}
\cos~\xi = \cos~\theta ~\cos\left(\omega~(t-t_0) \right)
~\cos~\theta_s + \sin~\theta ~\sin~\theta_s
\end{equation}
with $\theta$ the latitude on the comet, $t$ (s) the time since the beginning of the computation, $t_0$ (s) the initial time of computation, $\theta_s$ the cometocentric latitude of the sub-solar point that takes into account the obliquity of the comet (see Marboeuf et al. 2012), and $\omega=\frac{2\pi} {P_r}$ where $P_r$ (s) is the rotation period of the nucleus.

\subsection{Surface ablation and dust grains ejection at the surface of the nucleus \label{dustejection}}

At the beginning of the computation, the grains are encased in water ice and the comet has a homogeneous physical composition. By approaching the sun, the temperature of the surface increases. H$_2$O ice begins then to sublimate and grains can then be freed. At this time, the variation of the radius $\Delta R_n^{\phi, \theta}$ of the nucleus at the latitude $\theta$ and longitude $\varphi$ is then recomputed by using the following equation:
\begin{equation}
\Delta R_n^{\phi, \theta} = \sum \frac{M_x \Phi_x^{\varphi,
    \theta}}{\rho_{\varphi, \theta}} \Delta t \qquad (m)
\end{equation}
where $\Phi_x^{\varphi, \theta}$ (kg m$^{-2}$ s$^{-1}$) and $\rho_{\varphi, \theta}$ (kg m$^{-3}$) are respectively the flow of the gas $x$ and bulk density of ice and dust at the position ($\varphi$, $\theta$) on the surface of the nucleus.  Grains freed by sublimation of H$_2$O ice can either be ejected from the nucleus or accumulate at its surface thus forming a dust mantle covering the icy dusty layers. The formation of a dust mantle on the surface of the nucleus is mainly a function of the size of
grains initially embedded in the water ice and flow of gas escaping from the nucleus at a given latitude and longitude ($\varphi$, $\theta$). The largest radius $r^*$ of dust grain that can be ejected from the elemental surface ($\varphi$, $\theta$) of the nucleus is computed by comparing the sum of gas drag and centrifugal force with the gravitational attraction of the nucleus (see Orosei et al. 1999; Prialnik et al. 2004; Huebner et al. 2006; Marboeuf et al. 2012):
\begin{equation}
r^*_{\varphi, \theta}= \frac{3}{4} \frac{\sum_x M_x \Phi_x^{\phi,
    \theta} V_x}{\rho^{b,~d} (G_c \frac{M_n}{R_n^2} - R_n \omega^2
  cos^2\theta)} \qquad (m)
\label{Eq:largestdustgrains}
\end{equation}
\noindent where $V_x$ (m s$^{-1}$) is the velocity of the gas $x$, $\rho^{b, d}$ is the bulk density of the non porous dust grain (kg m$^{-3}$), $G_c$ (m$^3$ kg$^{-1}$ s$^{-2}$) is the gravitational constant, and $M_n$ (kg) is mass of the cometary nucleus. In this model, no cohesive forces between grains (see description in Huebner et al. 2006) are taken into account.  Note that dust grains with radii smaller than $r^*$ are immediately ejected to space, while larger
grains stay on the surface and contribute to the formation of a dust mantle.

Note that the size distribution of dust grains encased in H$_2$O ice is given by a power law (Rickman et al. 1990):
\begin{equation}
N(a) {\rm d}a= N_0 a^{\beta} {\rm d}a
\end{equation}
where $\beta$ is the power law index of the size distribution and $N_0$ a normalization factor.

\subsection{Density and porosity changes}
At the end of each time step $\Delta t$, we calculate the variation of mass density of species $x$ in each layer $i$ and position ($\varphi$, $\theta$) using $Q_x$:
\begin{equation}
\Delta \rho_x^{\varphi, \theta, i} = M_x Q_x^{\varphi, \theta} \Delta t \qquad (kg \,
m^{-3})
\end{equation}

For a more realistic physical representation of the nucleus, the porosity and the radius of the pores are also recomputed:

\begin{equation}
\Psi_{\varphi, \theta, i}= 1- \sum_x {\frac{\rho_x^{\varphi, \theta}}{\rho^{b,~x}}} 
\end{equation}

\begin{equation}
r_p^{\varphi, \theta, i}= r_p^i \sqrt{\frac{\Psi_{\varphi, \theta, i}}{\Psi^i_{\varphi, \theta, i}}}
\end{equation}
where $\rho_x^{\varphi, \theta}$ and $\rho^{b, x}$ are respectively the mass density and bulk density of the solid phase of the component $x$, and $\Psi_{\varphi, \theta, i}$ and $r_p^{\varphi, \theta, i}$ the initial porosity and pore radius at the position position ($\varphi$, $\theta$, i) in the nucleus.

\subsection{Orbital position and time step}
At the end of each time step, the orbital position $r_h$ of the comet and time step $\Delta t$ of the computation are calculated as follow:
\begin{eqnarray}
r_h & = & a(1-e~\cos{x_i}) \qquad (AU) \\
\Delta t & = &\frac{a^3}{G M_s} (1 - e~\cos{x_i})~ \Delta x_i \qquad (s)
\end{eqnarray}
where $a$ (AU) is the semi-major axis of the orbit of the comet, $e$ its eccentricity, $x_i$ the eccentric anomaly, $M_s$ the mass of the sun and $\Delta x_i$ the angular step.  At the beginning of the computation (at aphelion), $x_i=-\pi$, $r_h=a(1+e)$, $\Delta t=\Delta t_0$ (which is a fraction of a day), and $\Delta x_i =\frac{\Delta t_0}{\frac{a^3}{G M_s} (1 + e)}$.  As the comet approaches the sun, $r_h$ and $\Delta t$ decrease.

\section{Parameters \label{paramsection}}

The orbital and physical parameters adopted in this study are those of the comets 67P/Churyumov-Gerasimenko and 153P/Ikeya-Zhang given in Table \ref{paramss}. The comet's radii (2~km) and rotational period (12.76~hr) adopted for the two nuclei are those of the comet 67P/Churyumov-Gerasimenko (Lowry et al. 2012).

\paragraph{Chemical composition and water ice structure}
At the beginning of the computation, the cometary nucleus has a homogeneous composition of ices and dust. The ice phase of comets is mostly composed of H$_2$O, CO, CO$_2$, CH$_4$ and H$_2$S  volatile molecules. These molecules are the most abundant volatile species (production rates relative to water larger than 1\%) observed in cometary comae (Bockel\'ee-Morvan et al. 2004) with known equilibrium pressure curves of single guest clathrates.
Other equally abundant cometary volatile molecules such as H$_2$CO, CH$_3$OH and NH$_3$ are not considered in this study for the following reasons. H$_2$CO is not produced at the surface of the nucleus but is rather the result of a distributed source in the coma (Fray et al. 2004; Fray et al. 2006; Cottin \& Fray 2008). To our best knowledge, no experimental data concerning the stability curve of the CH$_3$OH clathrate has been reported in the literature (Marboeuf et al. 2008) and the conditions under which it forms stoichiometric hydrates or clathrate (Blake et al. 1991; Notesco \& Bar-Nun 2000) is still unclear. Finally, NH$_3$ does not form clathrate, but rather stoichiometric hemihydrates (2NH$_3$$-$H$_2$O) and/or monohydrates (NH$_3$$-$H$_2$O) under some conditions (Lewis 1972; Bertie \& Shehata 1984; Lunine \& Stevenson 1987; Kargel 1998; Moore et al. 2007) and is likely to be released together with H$_2$O, or only at slightly lower temperature (Schmitt et al. 1988). 

In this study, we consider four models of nuclei each with a different H$_2$O ice structure: amorphous, crystalline, clathrate and a mixture of these three structures (called hereafter `mixed' model). Hereafter, the amorphous model is named `nominal' model because it is the historical reference in cometary papers (Espinasse et al. 1991; De Sanctis et al. 2005; Prialnik et al. 2008; Rosenberg \& Prialnik 2009).
The others volatile molecules CO, CO$_2$, CH$_4$ and H$_2$S are condensed as pure ices in the porous network and/or trapped in amorphous water ice and/or in the clathrate structure. Their initial distribution between these states strongly depends on the environment temperature and pressure, the molecule (equilibrium pressure, size, polarizability) and their initial abundance in the molecular cloud (Kouchi et al. 1994; Bar-Nun et al. 2007) or in the protoplanetary disc, leading to a very large diversity among trapping efficiencies of various gases in amorphous water ice (Bar-Nun et al. 2007, Yokochi et al. 2012) and/or clathrate hydrates. 
Up to now no general agreement exists on the respective amounts of trapped gas (in amorphous water ice and/or clathrates) and gas condensed as pure ices in comets. Only some constraints on the trapping efficiency (maximum amount of species trapped) have been obtained experimentally.
Schmitt et al. (1989) showed that amorphous ice can trap other volatile molecules only up to a total of 8\% (in mole, relative to water) until crystallization occurs (Schmitt et al. 1992). The clathrate hydrate structure can trap up to 1/$n_{hyd}$ volatile molecules (about 17\%), where $n_{hyd}$ is the hydrate number defined as the molecular ratio $\frac{H_2O}{gas}$ in the clathrate. We fix hereafter the hydrate number $n_{hyd}$ to 6, closer to real values (see Marboeuf et al. 2012).

Table \ref{param_ices} gives the initial x/H$_2$O ($J_x$) mole fractions of the species $x$ ($x=$ CO, CO$_2$, CH$_4$ or H$_2$S) condensed either as pure ices in the porous network or trapped in the water ice following its structure (amorphous\footnote{No volatile molecule is trapped in crystalline ice since no experiment has shown this possibility. \label{footnote2}} or clathrate). Since the amount of trapped species and the process of trapping is different for the different water ice structures, it is natural to have nuclei that don't have the same abundances and chemical composition between the crystalline, amorphous and clathrate models. However, the values of $J_{\rm X}$ (sum of the three states) of all models are consistent with the observations in cometary comae of molecular species that are released directly from the nucleus (Bockel\'ee-Morvan et al. 2004).
The abundance of trapped species in amorphous water ice depends on many parameters.
As no reliable experimental data exists yet on the composition of mixed gas trapped in amorphous ice, we therefore choose, for the amorphous model, one nominal arbitrary set of plausible distribution of the volatile molecules between the states `trapped' inside amorphous water ice (up to a total of 8\% in mole, Schmitt et al. 1992) and `condensed' as pure ice (thus segregated from water ice) in the porous network, considering the relative equilibrium pressures of species at very low temperature (hereafter 30 K) and our current knowledge on trapping processes. 
For the crystalline model, the volatile molecules are only in the state `condensed' as pure ice. 
For the clathrate model, the distribution between the states `trapped' inside the clathrate structure (up to about 17\% in mole) and `condensed' as pure ice come from Mousis et al.~(2010). CO, CH$_4$ and H$_2$S represent respectively 79\%, 10\% and 11\% of the molecules trapped in the clathrate structure.
The "mixed" model is arbitrarily composed of 33\% of each of the three water ice structures presented before.

\paragraph{Ice to dust mass ratio}
The dust/ice mass ratio $J_{dust}$ is assumed to be equal to 1. This is approximately the value indicated for comet 1P/Halley by Giotto-DIDSY measurements (McDonnell et al. 1987), prescribed by Greenberg's (1982) interstellar dust model (Tancredi et al. 1994) and given by Lodders (2003) for solar system and photospheric compositions. It is also the most adopted value in cometary models (see Kossacki et al. 1999; Cohen et al. 2003; De sanctis et al. 2005; Lasue et al. 2008; Prialnik et al. 2008).
The size distribution of dust grains in comets follows a power law of order $-3.5$ (McDonnell et al. 1986; Huebner et al. 2006) with a upper cut-off at a radius of 1 cm (Prialnik 1997). The initial temperature is assumed to be equal at 30 K.

\begin{table*}
\centering 
\caption{Initial parameters for the cometary nucleus.}
\begin{tabular}{llcc|c}
\hline
Parameter		& Name & Units									& \multicolumn {2} {c}{Value}				\\ 
\hline
\multicolumn {4} {l}{Orbital parameters}    \\
67P/Churyumov-Gerasimenko\\
$a$ & semi-major axis				&		AU			&   \multicolumn {2} {c}{3.511}			\\
$e$ & eccentricity					&	  			&	  \multicolumn {2} {c}{0.632}  \\	
$q$ & perihelion            &   AU    &   \multicolumn {2} {c}{1.29}  \\
$R$ & Nucleus radius  & km 									& \multicolumn {2} {c}{2$^a$} 	 		\\
$P_r$ & Rotational period & h						& \multicolumn {2} {c}{12.76$^a$} 				\\
$P_o$ & Orbital period    & year      & \multicolumn {2} {c}{6.6} \\
      & Obliquity & degree &    \multicolumn {2} {c}{0} \\
153P/Ikeya-Zhang\\
$a$ & semi-major axis				&		AU			&   \multicolumn {2} {c}{50.0}			\\
$e$ & eccentricity					&	  			&	  \multicolumn {2} {c}{0.99}  \\
$q$ & perihelion            &   AU    &   \multicolumn {2} {c}{0.5}  \\	
$R$ & Nucleus Radius  & km 									& \multicolumn {2} {c}{2} 	 		\\
$P_r$ & Rotational period & h						& \multicolumn {2} {c}{12.76$^a$} 				\\
$P_o$ & Orbital period    & year      & \multicolumn {2} {c}{366} \\
      & Obliquity & degree &    \multicolumn {2} {c}{0} \\
\hline
\multicolumn {4} {l}{Physical parameters}    \\
$\Psi^i$  & Initial porosity &    \% &		\multicolumn {2} {c}{70}			\\
$\rho$    & Initial density    & kg m$^{-3}$ & \multicolumn {2} {c}{434} \\ 
h  &	Hertz factor    &      &    10$^{-2}$  & 10$^{-1}$  \\
K  &        Heat conductivity at perihelion & W m$^{-1}$ K$^{-1}$   &     $\approx$ 5 10$^{-3}$  & $\approx$ 5 10$^{-2}$\\
   &  Thermal inertia at perihelion  &    W m$^{-2}$ K$^{-1}$ s$^{1/2}$   &     $\approx$ 30 &  $\approx$ 90  \\
$T$ & Initial temperature & K 							&  \multicolumn {2} {c}{30} 				\\
$\epsilon$ & Infrared surface emissivity  & 									& \multicolumn {2} {c}{1}					\\
$A_l$ & Bolometric Bond's Albedo  & 										&  \multicolumn {2} {c}{0.05}			\\
$\tau$ & Tortuosity &   							& \multicolumn {2} {c}{$\sqrt{2}$$^{(b, c)}$}		\\
$r_p$  & Average pore radius 	&  m						& \multicolumn {2} {c}{10$^{-4}$} 				\\
$\beta$ & Power law size distribution of grains &  & \multicolumn {2} {c}{-3.5$^{(d)}$}	\\
 & Range size of grains &   m & \multicolumn {2} {c}{10$^{-6}$-10$^{-2}$}  \\ 
$J_{dust}$ & Dust/ice mass ratio & 					&  \multicolumn {2} {c}{1}  \\ 
\hline
\end{tabular}
\label{paramss}
\begin{flushleft}
$^{(a)}$Lowry et al. (2012); $^{(b)}$Kossacki \& Szutowicz (2006); $^{(c)}$Carman (1956), Mekler et al. (1990); $^{(d)}$ McDonnell et al. (1986), Huebner et al. (2006)
\end{flushleft}
\end{table*}

\begin{table*}
\centering 
\caption{Distribution of species in models.}
\begin{tabular}{l|c|cc|cc|ccc}
\hline
Models		& Crystalline	& \multicolumn {2} {c|}{Amorphous (`nominal')} &  \multicolumn {2} {c|}{Clathrate} &  \multicolumn {3} {c}{Mixed}				\\ 
\hline
Molecules$^1$ & condensed & condensed&    trapped                  &  condensed & trapped &    condensed & trap. amorph. & trap. clath.  \\
\hline
CO			&  10       &  7   & 3        									&     1   & 13   				& 10   & 1.3  & 4.4  \\ 
CO$_2$   &   5       & 2   & 3         									&     5   & 0    				& 4    & 1    & 0 \\
CH$_4$   &    2      & 1.2 & 0.8       									&   1.2 & 1.6  					& 1.2  & 0.34  & 0.56 \\  
H$_2$S   &  2        & 0.8  & 1.2      									&     1   & 1.8  				& 1    & 0.33  & 0.62 \\ 
\hline
Total    &  19       & 11   & 8                         &   8.2   &  16.4       & 16.2  & 2.97 & 5.58 \\
\hline
Total Model & 19     & \multicolumn {2} {c|}{19}        & \multicolumn {2} {c|}{24.6} & \multicolumn {3} {c}{24.75} \\
\hline
\end{tabular}
\begin{flushleft}
$^1$ Values are in \% of moles relative to H$_2$O
\end{flushleft}
\label{param_ices}
\end{table*}

\paragraph{Density, thermal inertia and thermodynamic parameters}

The density of cometary nuclei, obtained from observations and space missions, is estimated to range from 100 to 1000 kg m$^{-3}$ (Davidsson et al. 2009; Richardson et al. 2007; Lamy et al. 2007; Davidsson $\&$ Gutierrez 2005; Weissman et al. 2004; Davidsson \& Skorov 2002; Asphaug \& Benz 1994, 1996). These values, coupled with nucleus size measurements, lead to porosities greater than 50\%. In this work, we have chosen a porosity of 70\% leading to a bulk density of about 430 kg m$^{-3}$. This density is in good agreement with the range of values estimated for comet 67P/Churyumov-Gerasimenko by Davidsson \& Guti\'erriez (2005). 

An important cometary parameter is the thermal inertia, i.e. the thermal conductivity $K^m$ (by using the relation $I=\sqrt{\rho c K^m}$) of the cometary material. The thermal inertia of the upper layers of a comet nucleus can vary by several orders of magnitude from 40 to 3000 W m$^{-2}$ K$^{-1}$ s$^{\frac{1}{2}}$ (Davidsson et al. 2009) depending of the type of surface. Recently, Davidsson et al. (2013) have shown that some terrains of the dusty surface of the 9P/Tempel 1 comet have thermal inertia that can vary from less than 50 to about 150 W m$^{-2}$ K$^{-1}$ s$^{\frac{1}{2}}$ in one day. A low/high thermal inertia increases/decreases the temperature of the surface of the nucleus, leading respectively to an increase/decrease in the rate of sublimation of near surface ices and in the mass loss rate (see Marboeuf et al. 2012). The rate of production of gases from the nucleus could therefore be quite different between the extremes values. Moreover, it should be noted that these thermal inertias have been estimated assuming dusty surfaces without ices, i.e. for the dust crust. The thermal inertia of deeper porous icy cometary materials considering the sublimation of ices remains unknown today.
In this work, we test values of the Hertz factor $h$ of 10$^{-1}$ and 10$^{-2}$ W m$^{-1}$ K$^{-1}$, leading to thermal inertias of about 90 and 30 W m$^{-2}$ K$^{-1}$ s$^{1/2}$ respectively at perihelion.

Nuclei could build globally or locally a permanent dusty mantle at their surfaces that could also change their thermal behavior and the outgassing profiles and rates of production of species.
Current estimates from cometary models of the thickness of such a dust layer lies in the 1-10 cm range for 67P/Churyumov-Gerasimenko (Prialnik et al. 2008; Rosenberg \& Prialnik 2009), but are very uncertain since they are a function of many parameters such as orbit, dynamical history, gravity, size of grains, ect.
Given this expected range of thicknesses we will study the effects of the presence of a thin permanent dust mantle at the surface of the nucleus, varying the thickness from 1 to 50 cm.

The other thermodynamics parameters such as equilibrium pressures, enthalpies of sublimation, bulk densities of cometary materials (ices and dust), bulk thermal conductivities and heat capacities of solids are given in Marboeuf et al. (2012).

In this paper, we study the four models with different water ice structures and distributions of species between the `condensed' and `trapped' states with a low thermal inertia (30 W m$^{-2}$ K$^{-1}$ s$^{1/2}$). The chemical composition and distribution of species are given in Table \ref{param_ices}.\\
As the water ice structure is not the only parameter that induces changes in the outgassing profiles of volatile species, we study with the `condensed' and `trapped'nominal' model (amorphous) the influence of some physical parameters such as the thermal inertia, the presence of a permanent dust layer, and the abundance and distribution of species between the `condensed' and `trapped' states on the outgassing behavior of species and their relative abundances (to H$_2$O) near the surface of comets.
Table \ref{param_ices2} provides the abundances and distribution of species in the nucleus for the `nominal' model as well as for `composition 2', and `distribution 2' models. The `composition 2' model simulates a comet nucleus strongly enriched in CO$_2$ and impoverished in CO with a CO$_2$/CO ratio 4 times larger than the one of the `nominal' model. The `distribution 2' model simulates a more/less efficient trapping of the less/highly volatile species CO$_2$ and H$_2$S/CO and CH$_4$ in amorphous ice during the formation of the cometary material. 
For thermal inertia we test a `high value' case at 90 W m$^{-2}$ K$^{-1}$ s$^{1/2}$. For the dust mantle at the surface of the nucleus we test a global permanent layer ranging from 0.05 to 0.50 m thick with a thermal inertia of 42 W m$^{-2}$ K$^{-1}$ s$^{1/2}$. Note that the `high inertia' and `dust mantle' models have the same chemical composition and distribution of species between the `trapped' and `condensed' states as the `nominal' model.

\begin{table}
\centering 
\caption{Distribution of species in amorphous models.}
\begin{tabular}{l|cc|cc|cc}
\hline
Models		& \multicolumn {2} {c|}{nominal$^1$} & \multicolumn {2} {c|}{distribution 2} &  \multicolumn {2} {c}{composition 2}				\\ 
\hline
Molecules$^2$ &   cond. & trap. &   cond. & trap. &    cond. & trap.  \\
\hline
CO				&    7& 3 						&	 9   & 1   				& 4   & 1   \\ 
CO$_2$   &     2 & 3 						& 1   & 4    				& 5    & 5    \\
CH$_4$    &    1.2  & 0.8       & 1.8 & 0.2  				& 1  & 0.5 \\  
H$_2$S   &     0.8  & 1.2       &  0.5   & 1.5  		& 0.5    & 1.5   \\ 
\hline
Total    &  11   & 8            &   12.3 &  6.7      & 10.5  & 8 \\
\hline
Total Model & \multicolumn {2} {c|}{19}    & \multicolumn {2} {c|}{19}        & \multicolumn {2} {c}{18.5} \\
\hline
\end{tabular}
\begin{flushleft}
$^1$ The `high inertia' and `dust mantle' models have the same chemical composition and distribution of species between the `trapped' and `condensed' states as the `nominal' model.
\end{flushleft}
\begin{flushleft}
$^2$ Values are in \% of moles relative to H$_2$O.
\end{flushleft}
\label{param_ices2}
\end{table}

\section{Results \label{results}}

We study the outgassing behavior and the physical differentiation of the comet 67P/Churyumov-Gerasimenko by varying different physical parameters in the nucleus. Note that results are shown during one revolution and after 50 years of revolution around the sun (about 8 revolutions). However, they don't change significantly during this time scale, mainly because we present hereafter outgassing per unit of surface area of the nucleus in order to overcome the nucleus size which changes with time. \textbf{Note that we ran the models during 31 orbits (200 years) and that the results did not change significantly during this time scale.} Note \textbf{also} that except when it is mentioned, nuclei have no crust mantle at their surface.
In Sec.~\ref{sec:structure}, we first study and compare the outgassing behavior of volatile species (outgassing profiles along an orbit, day/night sides variations, ...), their differentiation in the nucleus (sub-surface structural and chemical stratigraphies, ...) and their relative abundance (to H$_2$O) in the coma for models made of fully crystalline ice, amorphous ice (the `nominal' model), clathrate or a mixture of the three. In a second study (Sec.~\ref{dustmantlemodels}), we study the influence of a thin dust mantle on the outgassing profile of species. We also vary the physical properties of the amorphous model such as the thermal inertia, the abundance of volatile molecules and their distribution between the `condensed' and `trapped' states (Sec.~\ref{physicalproperties}). These results are compared to the `nominal' (amorphous) model. 
Finally, we compare all the results of these models to observational data of relative (to H$_2$O) abundances of CO and CO$_2$ of several short and long period comets (Sec.~\ref{observations}).

\subsection{Influence of the water ice structure and volatility of species on their relative outgassing abundances \label{sec:structure}}

\paragraph{Physical differentiation of the nucleus}

\begin{figure*}
\begin{center}
\includegraphics[width=15.cm]{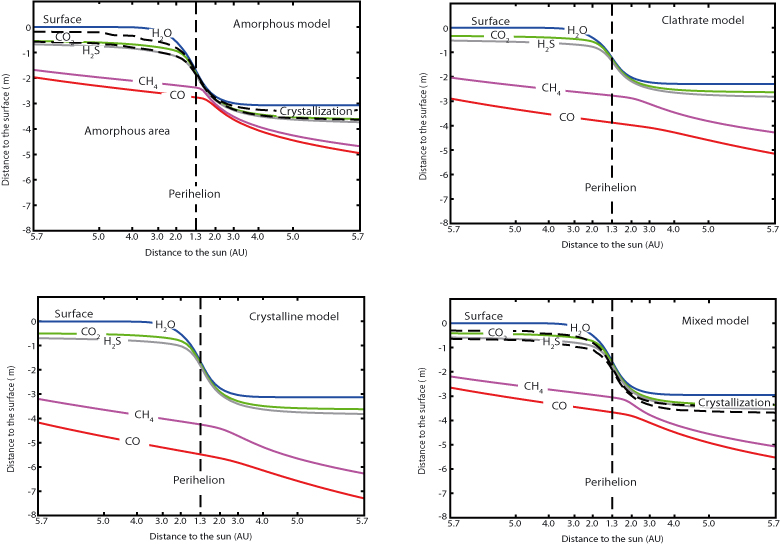}
\caption{\textbf{H$_2$O ice structure models} - Physical differentiation of the nucleus for all models (amorphous, crystalline, clathrate and mixed) as a function of the distance to the sun, during one revolution (the horizontal axis is linear in time). Calculations have been performed for the latitude $\theta$ = 10$^\circ$ (equator is 0$^\circ$). Thermal inertia $\approx$ 30 W m$^{-2}$ K$^{-1}$ s$^{\frac{1}{2}}$. The vertical dashed line corresponds to perihelion.}
\label{fig11}
\end{center}
\end{figure*}

\begin{figure*}
\begin{center}
\includegraphics[width=15.cm]{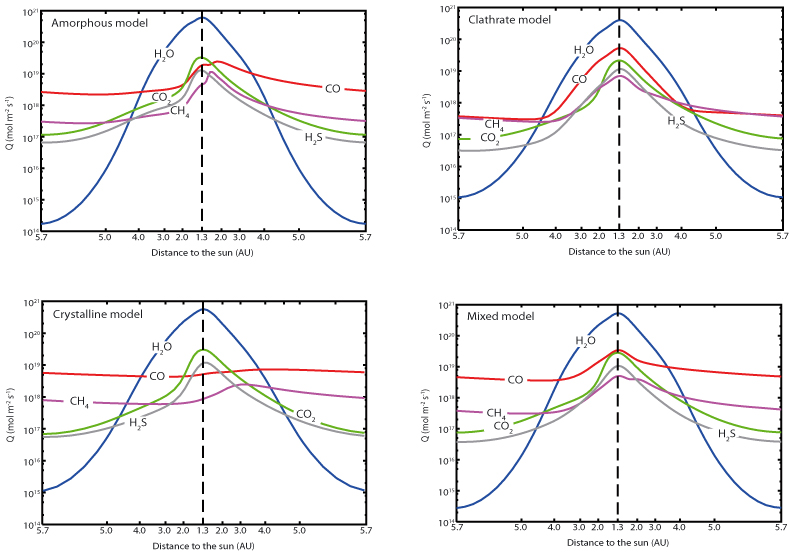}
\caption{\textbf{H$_2$O ice structure models} - Outgassing rate profiles, in molecules per second and per unit of surface area, of volatile molecules H$_2$O, CO, CO$_2$, CH$_4$, and H$_2$S for all models (amorphous, crystalline, clathrate and mixed) as a function of the distance to the sun, during one revolution (the horizontal axis is linear in time). Thermal inertia $\approx$ 30 W m$^{-2}$ K$^{-1}$ s$^{\frac{1}{2}}$. The vertical dashed line corresponds to perihelion.}
\label{fig12}
\end{center}
\end{figure*}

Figure \ref{fig11} represents the thermodynamic evolution of the nucleus of 67P/C-G as a function of the distance to the sun, during one revolution, for all models. The lines represent, from the top to the bottom, the nucleus surface (composed of H$_2$O and dust grains) and the minimum depths at which solid CO$_2$, H$_2$S, CH$_4$ and CO exist (hereafter called the sublimation interfaces of pure ices). Below, the gas phase of a given molecule is in equilibrium with its pure condensed phase in the porous network. Above, only the gas phase exists in the porous network. For the amorphous model, the crystallization interface (dashed lines) \textbf{with its associated released of trapped species} occurs at a depth of less than 0.5 meter below the nucleus surface and is at most 0.5 m thick. Above, the H$_2$O ice structure is fully crystalline. Below, it is fully amorphous.
The depths of the sublimation interfaces of pure ices are a function of the volatility and abundance of the species, and the thermal inertia of the cometary material. The more volatile is the specie, the deeper is its sublimation interface.
The low thermal inertia assumed for these models induces a limited physical differentiation of the nucleus within a depth of only 8 meters for the crystalline model and less (5-6 m) for the other models. The stronger differentiation for the crystalline model is due to the higher thermal conductivity of the crystalline water ice structure relative to the amorphous and clathrate structures (see Marboeuf et al. 2012). At each perihelion passage, the average ablation of the surface (from 2 to 3 m) reaches the interfaces of sublimation of some pure ices and of crystallization (for amorphous and mixed models only). \textbf{This leads to an increase of the temperature of interfaces and to} a rapid crystallization of amorphous ice with its associated release of trapped species, as well as a strong sublimation of condensed volatile molecules. After each passage, shortly after perihelion, the comet nucleus is like new with a depth of only 0.5-2.5 m of differentiated ices.
It should be noted that the ablation of the surface (sublimation of H$_2$O ice and release of trapped dust grains) reaches the interfaces of sublimation of pure CO$_2$ and H$_2$S ices for all models, leading to a strong sublimation \textbf{(outgassing rate)} of these molecules (see Fig.~\ref{fig12}). The interface of sublimation of the highly volatile species CO and CH$_4$ are reached only for the amorphous model, leading in this case to a strong sublimation \textbf{(outgassing rate)} of the corresponding pure ices (see Fig.~\ref{fig12}).

\paragraph{Outgassing behavior of comets as a function of the water ice structure}

Figure \ref{fig12} presents the outgassing profiles of H$_2$O, CO, CO$_2$, CH$_4$ and H$_2$S, in molecules per second and per unit of surface area, of the comet nucleus as a function of the distance to the sun, during one revolution, and for the four models. Whatever the initial water ice structure, the H$_2$O production does not change significantly since it is mainly crystalline at the surface \textbf{as shown by figure \ref{fig11} and whatever the initial structure of water ice in the nucleus}. It is not the case for the other volatile molecules CO, CH$_4$, CO$_2$, and H$_2$S, \textbf{deeper in the nucleus}, which have outgassing profiles very sensitive to the structure of water ice (respectively  the thermal conductivity of the nucleus, the depth of sublimation's interfaces, the distribution between the  'condensed' and 'trapped' states of species).

For the crystalline model, the maximum production rates in the outgassing profiles of CO$_2$ and H$_2$S are centered on perihelion (at about 1.3 AU) as they are for H$_2$O, mainly because their sublimation's interfaces are the nearest to the one of H$_2$O. 
For these two molecules, their profiles are not symmetrical about the perihelion passage. This results of the time required for the thermal wave from the surface to reach the depth where the sublimation fronts are located and increases with the species whose the depth of sublimation's interfaces are deeper:
the outgassing of CH$_4$ and CO increase slightly after perihelion with their maximum production shifted by about 2 and 3 AU respectively relative to perihelion. For all the molecules, the outgassing comes from the sublimation of the pure ices \textbf{initially} condensed in \textbf{the nucleus and which diffuses to the surface through the porous network.} The deeper the sublimation interface of the volatile molecule $X$ inside the nucleus (see Fig.~\ref{fig11}), the larger is the shift of the outgassing peak relative to perihelion and the lower is its amplitude.

For the amorphous model, the outgassing profiles of CO$_2$ and H$_2$S do not change markedly compared to the crystalline model: \textbf{as for the crystalline model, the maximum outgassing rates for these species are near perihelion. This maximum rate is due to the ablation of the surface which reaches the interfaces of sublimation of these species, and increases their rate of production.}
The only noticeable change \textbf{with the crystalline model} is the occurrence of an early low level production shoulder (about 1 order of magnitude weaker than the peak production) due partly to the release of trapped gases by \textbf{water} ice as \textbf{the crystallization of amorphous ice} starts before 5 AU (see Fig.~\ref{fig11}). This shoulder is also seen for CO and CH$_4$ but much weaker. For CO$_2$ and H$_2$S a larger contribution is added by the crystallization heat wave reaching the pure CO$_2$ and H$_2$S ice interfaces. However, the major observable changes are in the outgassing profiles of the highly volatile molecules CO and CH$_4$ which display a strong increase in production while approaching perihelion. This increase occurs in two steps. The first increase is due to the ablation of the surface that reaches the crystallization interface of amorphous water ice which induces \textbf{an increase of temperature and} a rapid crystallization \textbf{which releases} the trapped volatile molecules. The second increase occurs when the ablation reaches the sublimation interface of pure CO and CH$_4$ ices (see Fig.~\ref{fig11}), leading to \textbf{a strong increase of the temperature of these interfaces and the} sublimation of these molecules. 

For the clathrate model, remember that only CO, CH$_4$ and H$_2$S are trapped in the clathrate structure. CO$_2$ does not form a clathrate at low temperature and condenses only as a pure ice. In this case, only the outgassing profiles of the first three molecules are changed compared to the crystalline \textbf{and amorphous} model\textbf{s}. The maxima of outgassing of CO, CH$_4$ and H$_2$S all occur around the perihelion passage, in phase with the maximum of H$_2$O. \textbf{This maximum outgassing rate of the highly volatile species CO and CH$_4$} is due to the dissociation of the clathrate structure in surface and subsurface \textbf{layers which release the volatile species initially trapped inside:} \textbf{the interfaces of sublimation of these species are deeper in the nucleus relative to the surface compared to the amorphous model and then less affected by the thermal wave from the surface. The maximum production of H$_2$S is partly due to the dissociation of the clathrate structure and to the sublimation of the species since the ablation of the surface reaches the interface of sublimation as in the amorphous and crystalline models.} 

For the mixed model, \textbf{the maximum outgassing of all the molecules occurs near perihelion, as in the clathrate model, in phase with the maximum of H$_2$O.} 
Also the thermal behavior is close to that of the clathrate structure (see Fig.~\ref{fig11}). The outgassing profile of CO is more similar to the one of the clathrate model. \textbf{The major contribution comes from the dissociation of clathrates close-in subsurface layers with a contribution from the crystallization of amorphous water ice. However, because both contributions of dissociations of clathrates and crystallization of amorphous ice are strong in the outgassing rate of CO at perihelion passage, and because its interface of sublimation remains deeper than a few meters and then poorly affected by the thermal wave, there is no second maximum of CO production after perihelion passage such as in the amorphous model.} The outgassing of CH$_4$ is intermediate between the amorphous and clathrate models: \textbf{we observe a first increase in production at perihelion passage due to the dissociation of clathrates and the crystallisation  of amorphous water ice, and a second one after passage} due to the ablation of the surface which comes close to the interface of sublimation of its pure ice (see Fig.~\ref{fig11}). The outgassing profile of CO$_2$ and H$_2$S are very close to their corresponding profiles in the crystalline \textbf{and amorphous models} (although variations among models for CO$_2$ are not large): \textbf{the major contribution of these species come from the sublimation of their pure ices.}

\begin{figure*}
\begin{center}
\includegraphics[width=15.cm]{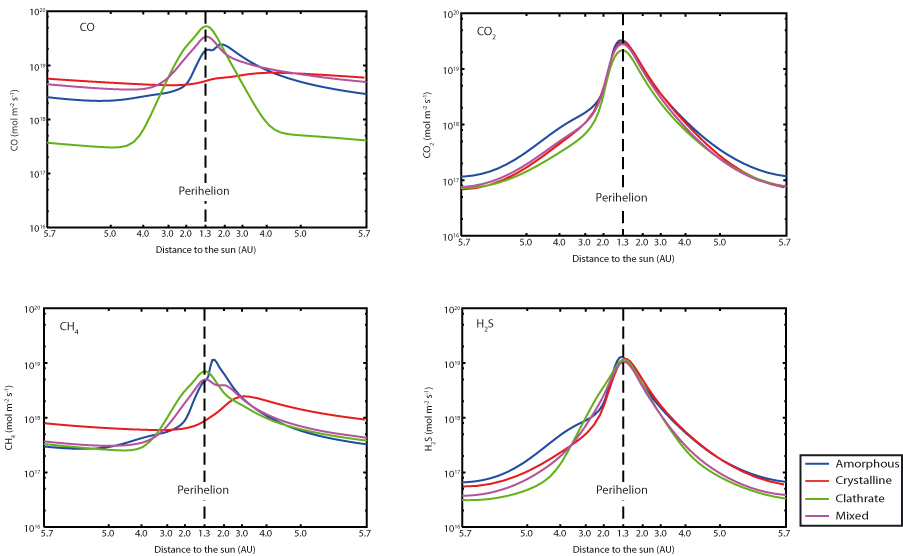}
\caption{\textbf{H$_2$O ice structure models} - Comparison of the outgassing profiles, in molecules per second and per unit of surface area, of the volatile molecules CO, CO$_2$, CH$_4$, and H$_2$S for all models (amorphous, crystalline, clathrate and mixed) as a function of the distance to the sun, during one revolution (the axis is linear in time). Thermal inertia $\approx$ 30 W m$^{-2}$ K$^{-1}$ s$^{\frac{1}{2}}$. The vertical dashed line corresponds to perihelion.}
\label{fig13}
\end{center}
\end{figure*}

Finally, the major observable difference between models (amorphous, crystalline, clathrate and mixed) comes from the outgassing profile of highly volatile molecules such as CO and CH$_4$.  
Figure \ref{fig13} presents the comparison of the outgassing profiles (in molecules per second and per unit of surface) of the volatile molecules CO, CO$_2$, CH$_4$, and H$_2$S, for all models. It is impossible to distinguish the water ice structure in the comet nucleus from the outgassing profile of less volatile molecules such as H$_2$O, CO$_2$, and H$_2$S. However, the highly volatile molecules CO and CH$_4$ allows one to distinguish the crystalline (relatively flat outgassing profile with a late post-perihelion increase) from the amorphous (one to two peaks of production: the first at perihelion and the second slightly later) and from the clathrate structures (one peak of production in phase with the one of H$_2$O, shortly preceded by a small pre-perihelion production shoulder). CO$_2$, and H$_2$S may also help to distinguish the amorphous structure against the others as they both, especially H$_2$S, have a wide and early low level pre-perihelion production shoulder, starting as early as 5 AU. CH$_4$ and CO also displays this shoulder, but much weaker.

\begin{figure*}
\begin{center}
\includegraphics[width=15.cm]{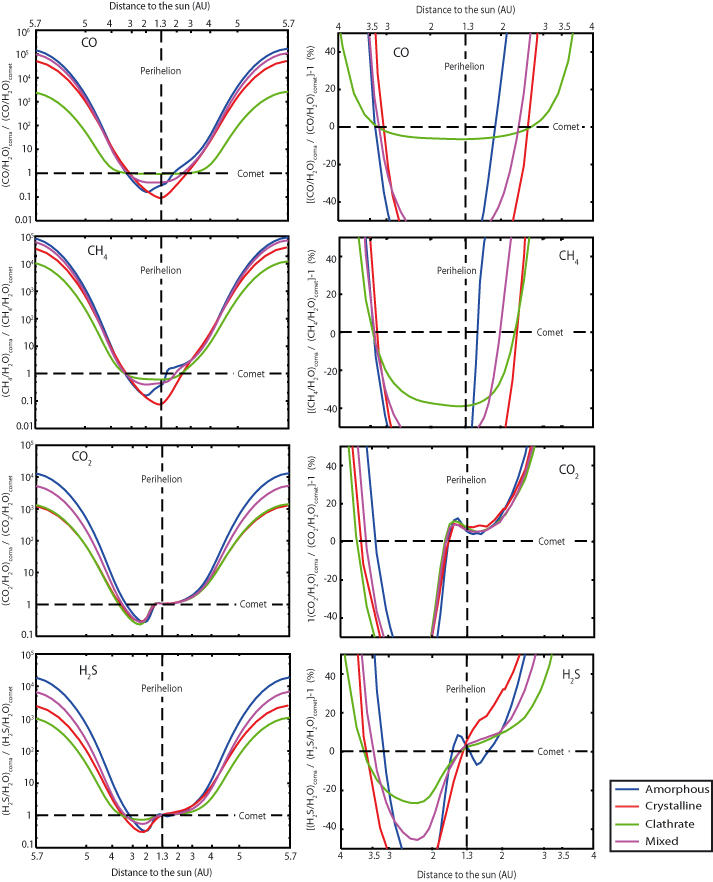}
\caption{\textbf{H$_2$O ice structure models} - Ratio X/H$_2$O of the gas productions in the coma relative to this ratio in the primitive nucleus (left column) and the deviation from the primitive composition of the nucleus (right column) for volatile species CO, CO$_2$, CH$_4$, and H$_2$S as a function of the distance to the sun, for all models (amorphous, crystalline, clathrate and mixed). The value 1 (horizontal dashed line, left column of the figure) corresponds to the primitive abundance. The value '0' (horizontal dashed line, right side of the figure) corresponds to the case where no deviation is observed between the coma and the nucleus. The vertical dashed line corresponds to perihelion. Thermal inertia $\approx$ 30 W m$^{-2}$ K$^{-1}$ s$^{\frac{1}{2}}$.}
\label{fig14}
\end{center}
\end{figure*}

\paragraph{Abundance of species in coma}
H$_2$O being the most abundant molecule in comets, all species are referred to this molecule for the abundances observed in the coma (see Bockel\'ee-Morvan et al. 2004; Dartois 2005; Mumma \& Charnley 2011).
Figure \ref{fig14} presents both the ratio $\frac{X}{H_2O}$ of the gas production rates in the coma relative to the initial ratio $\frac{X}{H_2O}$ in the solid phase (sum of all solid phases) of the nucleus (left column of the figure) and the deviation from the primitive composition of the nucleus (right column of the figure) for species CO, CO$_2$, CH$_4$, and H$_2$S as a function of the distance to the sun. The value `1' (horizontal dashed line, left column of the figure) corresponds to the normalized initial nucleus abundance for each species. The value '0' (horizontal dashed line, right column of the figure) corresponds to the case where no deviation is observed between the coma and the nucleus. The vertical dashed line corresponds to perihelion.
The relative abundances of all species vary by several orders of magnitude from aphelion (10$^3$-10$^5$) to perihelion (0.1-0.5)\textbf{, mainly due to the temperature of sublimation which vary greatly from one species to another (see Fray \& Schmitt 2009).}
Far away from perihelion, \textbf{where the surface temperature is low ($T_s\leq$100 K)}, the rate of H$_2$O production decreases significantly and results in a high relative abundance of the volatile species. Near perihelion passage, \textbf{where the surface temperature ($T_s\geq$190 K) increases significantly compared to aphelion ($T_s\leq$60 K),} the flux of H$_2$O gas increases strongly, leading to a significant decrease of the relative abundances of all the more volatile species.
Note that relative abundances of highly volatile molecules such as CO and CH$_4$ vary more than 1 order of magnitude in the range of heliocentric distances 3 to 1.3 to 3 AU near perihelion while less volatile molecules such as CO$_2$ and H$_2$S vary by less than 1 order of magnitude in the same range. 
 
For CO$_2$ and H$_2$S, their relative abundances decrease down to $\approx$ 0.5 before perihelion passage, and become approximately equal to the initial nucleus composition (horizontal dashed lines) from shortly before perihelion out to $\approx$ 2.4 AU, whatever is the model. The reason of this equality comes from the fact that the ablation of the surface (H$_2$O production) comes very close to the CO$_2$ and H$_2$S ice interfaces shortly before perihelion (see Fig.~\ref{fig11}) thus sublimating these volatile ices, either in the subsurface (trapped in amorphous ice and as pure ices) or at the surface (clathrates) at the same rate as surface ablation
 
For highly volatile molecules such as CO and CH$_4$, their relative abundances decrease down to about 0.1 around perihelion in the cases of the crystalline and amorphous models, while they remain approximately equal to the primitive nucleus composition (horizontal dashed lines) for the clathrate model and slightly lower for the mixed model. 
In fact the relative abundances are the result of the shift of maximum productions of the species relative to the one of H$_2$O (see Fig.~\ref{fig12}). With the clathrate model their relative perihelion productions closely correspond to the fraction trapped in the clathrate relative to the total initial amount of the species (93\% for CO, 57\% for CH$_4$, see Tab.~\ref{param_ices}) because the removal/ablation of clathrates at the surface controls the gas production with only minor contributions from pure ices sublimation (i.e. much less than their initial pure ice abundances: 7\% for CO, 43\% for CH$_4$, see Tab.~\ref{param_ices}) in the deep subsurface (see Fig.~\ref{fig11}). \textbf{For the amorphous and crystalline models, it is mainly the depth of interfaces of volatile species (relative to the one of H$_2$O), i.e. the sublimation of species and their diffusion through the porous network, which controls the gas production. For the amorphous model, the ablation of the surface reaches the crystallization area of the comet leading to an increase of the outgassing rate of highly species as described before. However, the fractions of volatile species initially trapped in amorphous ice (30\% and 40\% of the total CO and CH$_4$ in the nucleus, respectively) are always  smaller than in clathrates (93\% and 57\% of the total CO and CH$_4$ in the nucleus, respectively), and thus the outgassing rate of the amorphous model is always weaker than the one from the clathrate model.} 

So, only the models containing a significant amount of clathrates show relative abundances in the coma around perihelion in good agreement for ALL molecules with the initial ones in the nucleus. Amorphous and crystalline models present systematically lower coma abundances of the very volatile molecules, especially for the crystalline model, with up to 1 order of magnitude depletion.

However, for each model, there are two places on the orbit, shortly before and after perihelion, where all species are produced with a limited relative deviation ($\leq$ 50\%) from their initial nucleus abundances, although for CO and CH$_4$ they vary rapidly before and after these heliocentric distances. These pre and post heliocentric distances and the maximum deviations relative to the primitive composition are at -3.2 AU (25\%) and +1.5 AU (50\%) for the amorphous model, -3.4 AU (50\%) and +2.2 AU (50\%) for the crystalline model, -3.7 AU (25\%) and +2.1 AU (15\%) for the clathrate model, and -3.45 AU (25\%) and +2.1 AU (25\%) for the mixed one.
When considering all four models the smallest global maximum deviations for all species is 50\% at 3.5 AU pre-perihelion. For the less volatile molecules a much smaller deviation (less than 10\%) occurs for ALL models from about -1.8 AU (before perihelion) to about 1.8 AU (after perihelion passage) for CO$_2$ and -1.5 to 1.5 AU for H$_2$S.  
For the clathrate model there is even a much wider orbital range, from -2 AU to 2.5 AU where all species are produced at relative rates which depart by less than 40\% from the primitive nucleus composition (and less than 5\% for CO).

\begin{figure*}
\begin{center}
\includegraphics[width=15.cm]{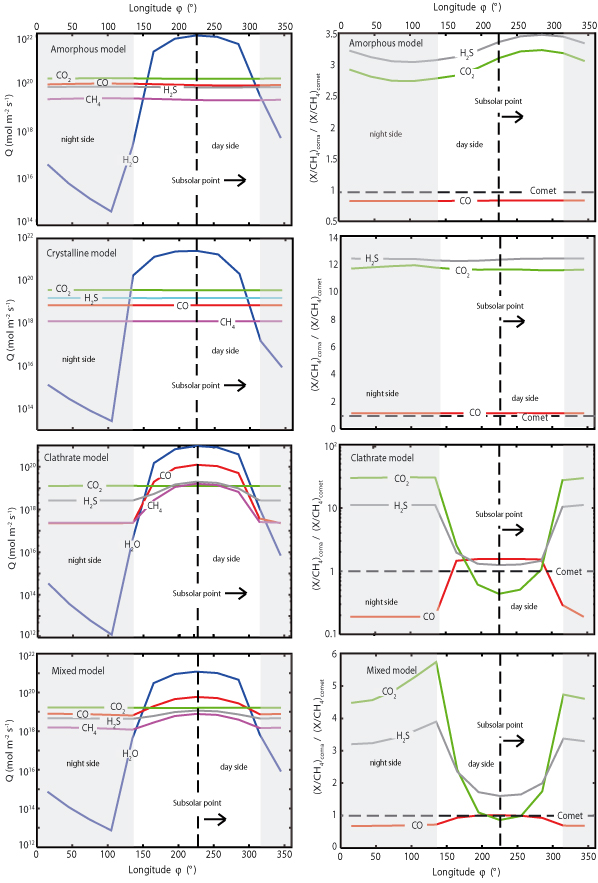}
\caption{\textbf{H$_2$O ice structure models} - Day/night variations of the outgassing rates, in molecules per second and per unit of surface area, of volatile molecules (left column) H$_2$O, CO, CO$_2$, CH$_4$, and H$_2$S for all models (amorphous, crystalline, clathrate and mixed), and day/night variation of the outgassing ratio X/CH$_4$ of volatile molecules CO, CO$_2$, and H$_2$S relative to the ratio in the nucleus (right column) as a function of the longitude $\varphi$ at perihelion passage. The noon meridian is indicated by a vertical dashed line and moves towards increasing longitudes as the nucleus rotates. Calculations have been performed for latitude $\theta$ = 10$^\circ$. Thermal inertia $\approx$ 30 W m$^{-2}$ K$^{-1}$ s$^{\frac{1}{2}}$.} 
\label{fig15}
\end{center}
\end{figure*}

\paragraph{Day/night variations of the outgassing}
The diurnal fluctuations of the gas productions are other indicators which can allow one to distinguish between the different structures of water ice in comets. Figure \ref{fig15} presents the day/night variations (day = hemisphere illuminated by the sun, night = hemisphere not illuminated by the sun) of the outgassing (in molecules per second and per unit of surface area) of volatile molecules H$_2$O, CO, CO$_2$, CH$_4$, and H$_2$S for all models (amorphous, crystalline, clathrate and mixed) and the ratios $\frac{X}{CH_4}$ of the gas productions in the coma relative to the initial ratio $\frac{X}{CH_4}$ in the solid phase of the nucleus for species CO, CO$_2$ and H$_2$S as a function of the longitude $\varphi$ at perihelion passage. Calculations have been performed for the latitude $\theta$ = 10$^\circ$. 

For the amorphous and crystalline models, only the outgassing of H$_2$O presents large variations with the diurnal illumination cycle of the rotating nucleus (see Fig.~\ref{fig15}). CO$_2$ and H$_2$S, trapped or/and condensed near the surface of the nucleus, show small diurnal variations (about 12\% in the morphous model, less than 3\% for the crystalline). While the more volatile molecules CO and CH$_4$, whose sublimation's interfaces are deeper \textbf{and less submited to the diurnal change temperatures}, are released steadily (see Fig.~\ref{fig15}) by the comet whatever the side (day or night).

In the case of the amorphous model, the crystallization of water ice appears to release volatile species with negligible influence of the day/night variations of insulation at the surface of the nucleus. The layer subjected to crystallization is quite close to the surface, 8 cm (see Fig.~\ref{fig11}), but remains below several diurnal thermal skin depths\footnote{The thermal skin depth is the distance over which the thermal energy decreases by a factor of 1/e}, which is about 1.5 cm for the low thermal conductivity considered here. However the less volatile molecules CO$_2$ and H$_2$S which have their pure ice interfaces at similar depth at perihelion, feel slightly the diurnal thermal wave, which modulates their sublimation by some 10\%, because their sublimation rates are more sensitive to small temperature variations at the perihelion subsurface temperature (118 K and 111K for CO$_2$ and H$_2$S respectively, at 10 cm and 15 cm below the surface respectively) than the ice crystallization rates.

For the amorphous model the relative gas fluxes escaping the nucleus (see Fig.~\ref{fig15}) show enhanced productions (by factors of 2.7 to 3.5) of the less volatile species CO$_2$ and H$_2$S (CO$_2$:H$_2$S:CH$_4$ $\approx$ 7.5:3.2:1) compared to their initial fraction in the nucleus (CO$_2$:H$_2$S:CH$_4$ $=$ 2.5:1:1, cf. Tab.~\ref{param_ices}). \textbf{This is mainly due to the high erosion rate of the surface which reaches the sublimation interface of the less volatile species while the interface of sublimation of CH$_4$, deeper in the nucleus, is not affected at perihelion.}
 \textbf{Also}, the highly volatile species CO is produced at a rate relative to CH$_4$ (CO:CH$_4$ = 4:1) quite close to its initial fraction in the nucleus (CO:CH$_4$ = 5:1) \textbf{because these species are released together from the crystallization front and diffuse from deeper interfaces of sublimation in the nucleus}.

For the crystalline model, the sublimation interfaces of the less volatile species CO$_2$ and H$_2$S are also quite close to the surface at about 8 cm depth (see Fig.~\ref{fig11}), but remain below several diurnal thermal skin depths which is of about 1.5 cm. The outgassing behavior of all species are then poorly affected by day/night variations of insulation. The relative gas fluxes of CO$_2$ and H$_2$S are strongly enhanced (CO$_2$:H$_2$S:CH$_4$ $\approx$ 30:12:1 in the coma) relative to their initial fraction in the nucleus (CO$_2$:H$_2$S:CH$_4$ $=$ 2.5:1:1, cf. Tab.~\ref{param_ices}). The gas flux of CO relative to CH$_4$ is approximately the same (only 10\% more) as in the primitive composition of the nucleus.
The enhanced production of the low volatility species compared to highly volatile molecules CO and CH$_4$ is due to the rapid ablation of the surface around perihelion (see Fig.~\ref{fig11}), where the less volatile species are condensed. Highly volatile species are condensed deeper in the nucleus and are affected later by erosion of the nucleus surface. That's why at perihelion the amorphous and crystalline models have relative (to H$_2$O) abundances of CO and CH$_4$ in coma less than the primitive composition in the nucleus (see Fig.~\ref{fig14}).

Only the clathrate and mixed models display diurnal variations of the outgassing of the species CO, H$_2$S and CH$_4$, in addition to H$_2$O. These molecules are released at the surface on the day side of the nucleus during the sublimation of the clathrate structure, and remains trapped (for CO, H$_2$S and CH$_4$) or condensed (H$_2$O) on the night side when the temperature decreases. For the clathrate model the day time fluxes of H$_2$S and CO species, relative to CH$_4$, are slightly enhanced (CO:H$_2$S:CH$_4$ $\approx$ 7.8:1.2:1 in coma) relative to their initial fraction in the nucleus (CO:H$_2$S:CH$_4$ $=$ 5:1:1, cf. Tab.~\ref{param_ices}). However, the day time flux closely corresponds to the fraction of species trapped in the clathrate (CO:H$_2$S:CH$_4$ $=$ 8.125:1.125:1, cf. Tab.~\ref{param_ices}) because on the day side the removal of clathrate at the surface controls the gas production rates with smaller contributions from sublimation of pure ices.

CO$_2$, not being trapped in clathrates at the surface, but condensed deeper, does not show variations, as for the crystalline and amorphous models. The night time fluxes of all the species mostly correspond to the sublimation of the condensed part of the volatile species with strongly enhanced relative productions of CO$_2$ and H$_2$S (CO$_2$:H$_2$S:CH$_4$ $\approx$ 55:10:1 in the coma) compared to their total initial fraction in the nucleus (CO$_2$:H$_2$S:CH$_4$: = 1.8:1:1, cf. Tab.~\ref{param_ices}), due to the rapid ablation of the nucleus surface around perihelion (see Fig.~\ref{fig11}).
Inversely, the relative production of CO in the coma (CO:CH$_4$= 1:1) decreases markedly compared to its initial total fraction in the nucleus (CO:CH$_4$ = 5:1, cf. Tab.~\ref{param_ices}) but corresponds more to the initial condensed fraction in the nucleus (CO:CH$_4$ = 0.83:1, cf. Tab.~\ref{param_ices}).

The difference in amplitude of the day side ougassing surge between the clathrate and mixed models comes from the fraction of clathrates in the forner (100\% of H$_2$O) compared to the latter (33 mole-\% of H$_2$O). The mixed model also presents lower enhanced productions of the species CO$_2$ and H$_2$S on the night side of the nucleus compared to the clathrate and crystalline models. For CO, the relative production is similar to the amorphous and crystalline models on the night side of the nucleus. For the day side of the nucleus, the relative production of species are quite close to the ones of the clathrate model.

The day/night variations, associated with the orbital outgassing profiles of volatile molecules, should allow one to help distinguish the clathrate structure from the other water ice structures in cometary nuclei.

\subsection{Dust mantle models \label{dustmantlemodels}}

In this section, we study the effects of a permanent dust mantle on the outgassing behavior of the nucleus, for several thicknesses (5, 10, 30 and 50 cm) of dust mantle. The `dust mantle' model simulates a more realistic nucleus with an accumulated homogeneous dust layer. Note that the thickness of the dust mantle does not change with time. We apply this study only to the amorphous model ('nominal' model hereafter); although the shape of the gas production curves depend on the water ice structure, the relative changes (rate, physical differentiation) are similar whatever the models. So, observations described in this section remain valid for all models whatever the water ice structure.

\paragraph{Physical differentiation of the nucleus}

\begin{figure*}
\begin{center}
\includegraphics[width=15.cm]{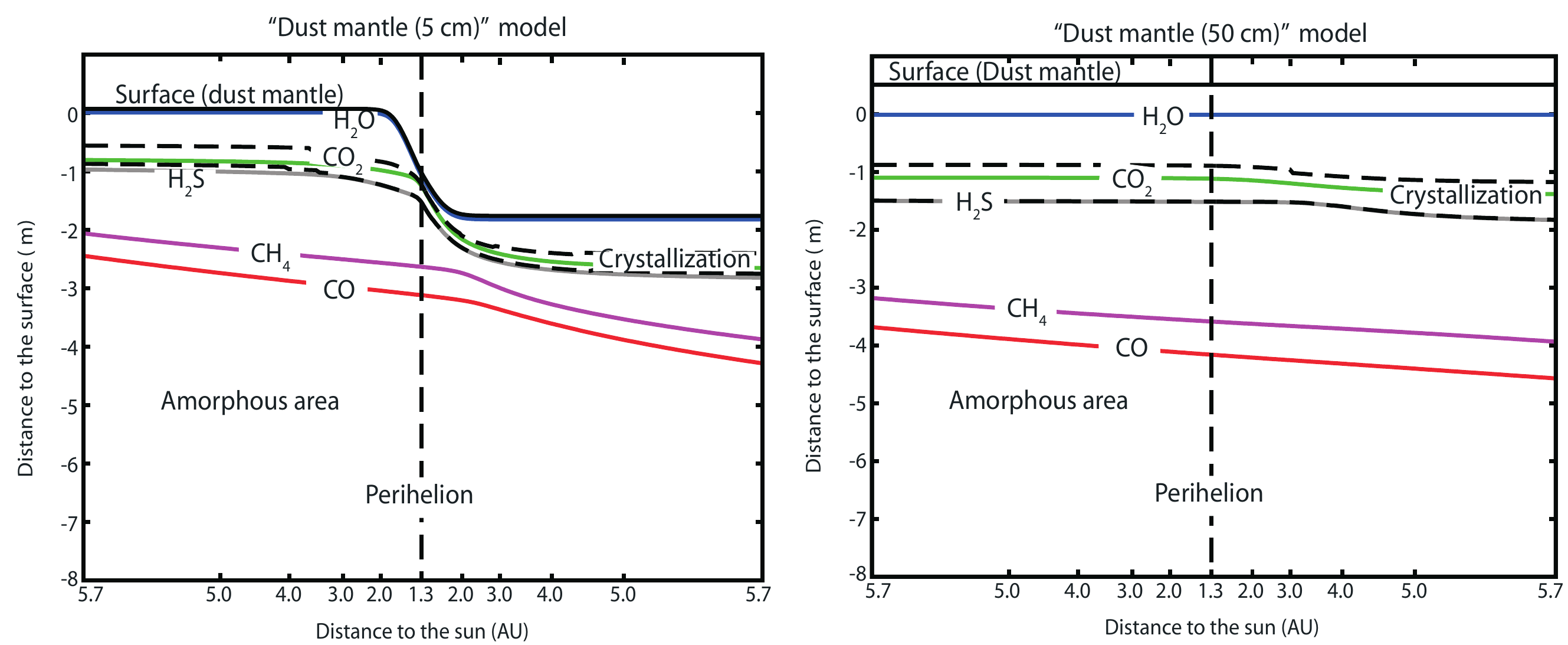}
\caption{\textbf{Dust mantle models} - Physical differentiation of the comet nucleus for dust mantle thicknesses of 5 and 50 cm at the surface of the nucleus as a function of the distance to the sun, during one revolution (the horizontal axis is linear in time). Calculations have been performed at latitude $\theta$ = 10$^\circ$. The thermal inertia $\approx$ 30 W m$^{-2}$ K$^{-1}$ s$^{\frac{1}{2}}$.}
\label{fig21}
\end{center}
\end{figure*}

Figure \ref{fig21} shows the physical differentiation of the nucleus as a function of the distance to the sun, during one revolution, for dust mantle thicknesses of 5 and 50 cm at the surface of the nucleus. The thermal inertia of the dust mantle is $\approx$ 41 W m$^{-2}$ K$^{-1}$ s$^{\frac{1}{2}}$. For comparison with the 'nominal' model, see Fig.\ref{fig11}.
For a 5 cm thick crust, the average ablation of the surface is limited to about 2 m per perihelion passage. The sublimation interfaces of all species and of crystallization, are deeper by some 10's of cm in the nucleus compared to the `nominal model', \textbf{due to the weaker erosion rate of the surface}. 
For a 50 cm thick crust model, the surface does not suffer erosion near perihelion passage (ablation $<$ 1 mm). The sublimation interfaces of all species and of crystallization, are deeper by some 10's of cm to 1 meter (especially near perihelion) in the nucleus compared to the `nominal model'. They are poorly affected at the perihelion passage with slow progression to the center of the nucleus at a rate of some 30 cm per revolution for CO$_2$ and H$_2S$ and almost 1 meter per revolution for CO and CH$_4$.

\paragraph{Outgassing behavior}

\begin{figure*}
\begin{center}
\includegraphics[width=15.cm]{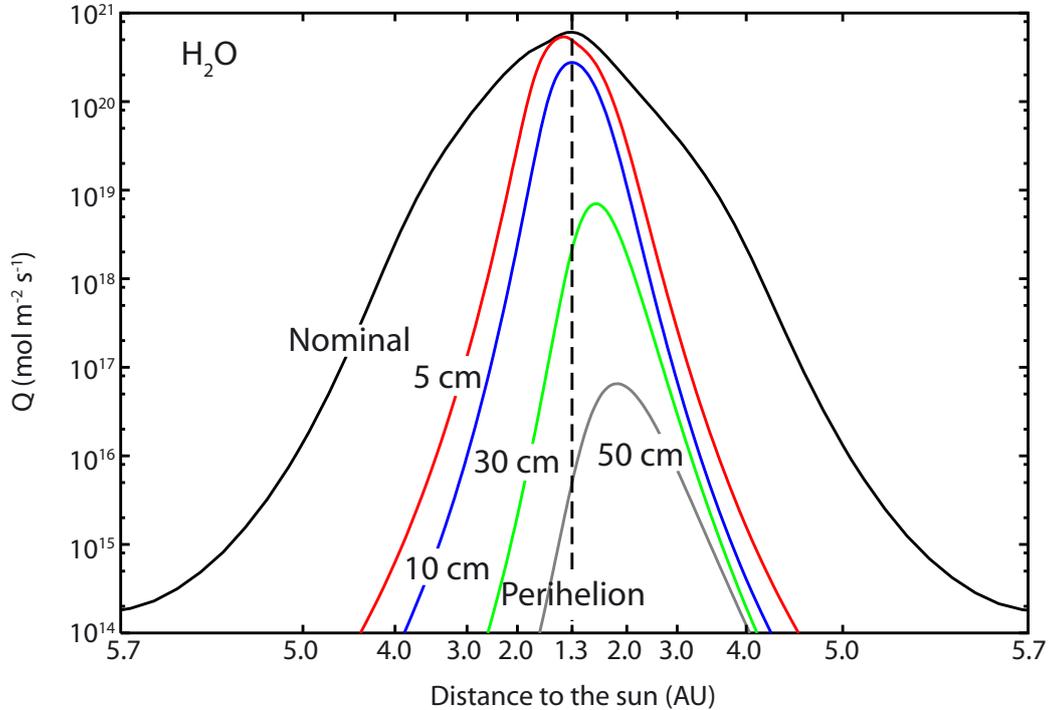}
\caption{\textbf{Dust mantle models} - Comparison of the outgassing profiles, in molecules per second and per unit of surface area, of H$_2$O for the `nominal' amorphous models without and with a dust mantle of different thicknesses as a function of the distance to the sun, during one revolution (the horizontal axis is linear in time). The vertical dashed line corresponds to perihelion.}
\label{fig22}
\end{center}
\end{figure*}

\begin{figure*}
\begin{center}
\includegraphics[width=15.cm]{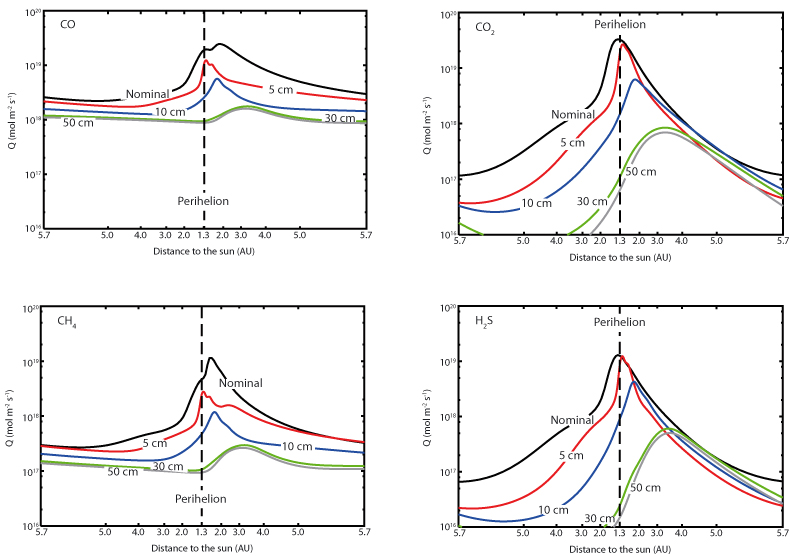}
\caption{\textbf{Dust mantle models} - Same as in Fig.~\ref{fig22} but for CO, CO$_2$, CH$_4$, and H$_2$S.}
\label{fig23}
\end{center}
\end{figure*}

Figures \ref{fig22} and \ref{fig23} present the comparison of outgassing profiles, in molecules per second and per unit of surface area, of H$_2$O, and CO, CO$_2$, CH$_4$ and H$_2$S respectively, as a function of the distance to the sun, during one revolution for all the 'dust mantle' models (crust thicknesses of 5, 10, 30 and 50 cm). For comparison, outgassing profiles of species for the 'nominal' model (black lines) are reported in the figures.

The major effect of the dust layer is to reduce the production rates of all species from about some 10's of percent to 1 to 2 orders of magnitude (up to 3 to 4 orders of magnitude for H$_2$O) compared to the `nominal' model and to shift the maximum production of species after perihelion passage.
The outgassing profiles of H$_2$O display a very steep, reduced and shifted (after perihelion) water production due to the insulating dust layer at the surface (lower temperature of the H$_2$O ice interface). The thicker is the crust, the weaker and more delayed is the water production. This results can be explained by the sublimation interface of H$_2$O that is deeper in the nucleus and affected later due to the time required for the thermal wave from the dusty surface to reach it. Whereas the dust model with a homogeneous dust layer of 5 cm shows a peak production of H$_2$O at perihelion of the same order of magnitude as the 'nominal' model, the dust model with a layer of 50 cm shows a peak of H$_2$O production shifted by 0.5 AU and weaker by 4 orders of magnitude.
\textbf{As H$_2$O,} the outgassing profiles of the other species CO, CO$_2$, CH$_4$ and H$_2$S are also markedly modified. 
For crust layers larger than 5 cm, all the species display a single peak of production shifted post-perihelion (see Fig.~\ref{fig23}). The presence of an insulating dust mantle at the surface of the nucleus results in a lower temperature of the water ice interface. The consequence is a reduction of the erosion and the physical differentiation of the nucleus (see Fig.~\ref{fig21}), and hence the rate of production of the volatile species from the nucleus (Figures \ref{fig22} and \ref{fig23}). The absence of the wide pre-perihelion production shoulder of CO$_2$ and H$_2$S for thicknesses greater than 5 cm, and of the narrower peak(s) or shoulder of CO and CH$_4$ around perihelion is due to the low amplitude of the crystallization events (see Fig.~\ref{fig21}, right column), \textbf{due to lower temperatures encountered in the layers of the nucleus}.

This set of results shows that extended observations (wide range of heliocentric distances) of the outgassing profiles of a series of species with different volatilities can help to constrain the presence of a dust mantle at the surface of the nucleus and its thickness.

\paragraph{Abundance of species in coma}

\begin{figure*}
\begin{center}
\includegraphics[width=15.cm]{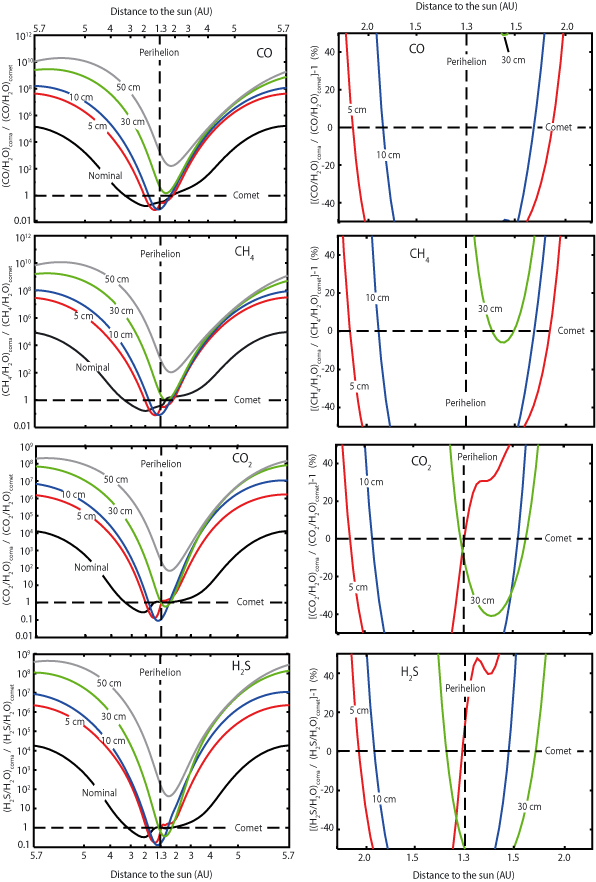}
\caption{\textbf{Dust mantle models} - Ratio X/H$_2$O of the gas productions in the coma relative to this ratio in the primitive nucleus (left column) and the deviation from the primitive composition of the nucleus (right column) for volatile species CO, CO$_2$, CH$_4$, and H$_2$S as a function of the distance to the sun, for the `nominal' amorphous models without and with dust mantle of different thicknesses. The value 1 (horizontal dashed line) corresponds to the primitive abundance. The vertical dashed line corresponds to perihelion.}
\label{fig24}
\end{center}
\end{figure*}

Figure \ref{fig24} presents both the ratio $\frac{X}{H_2O}$ of the gas productions in the coma relative to the initial ratio $\frac{X}{H_2O}$ in the solid phase (sum of all solid phases of X) of the nucleus (left column of the figure) and the deviation from the primitive composition of the nucleus (right column of the figure) for species CO, CO$_2$, CH$_4$, and H$_2$S as a function of the distance to the sun, for all 'dust mantle' models. The value `1' (horizontal dashed line, left column of the figure) corresponds to the normalized initial nucleus abundance for each species. The value '0' (horizontal dashed line, right column of the figure) corresponds to the case where no deviation is observed between the coma and the nucleus. The vertical dashed line corresponds to perihelion. For comparison with the 'nominal' model, the relative abundances for all species are also reported (black lines) on the figures.
Near aphelion, the production rate of all species increase relative to H$_2$O by 2.5 (crust thickness of 5 cm) to $\approx$ 5 orders of magnitude (crust thickness of 50 cm), mostly due to a decrease in the H$_2$O production rate in the same proportions (decrease of surface temperature, see Fig.~\ref{fig22}). Depending on the dust thickness the gas productions can vary between 10\% and 1000 times the primitive nucleus composition around perihelion.

For models with dust layers thinner than 10 cm there is only one place on the orbit where each species is produced with a relative deviation of less than 25\% from their initial nucleus abundances, although they vary rapidly before and after this heliocentric distance. This position varies from 3.2, to 2.1 to 1.85 AU pre-perihelion for the dust free model, and the 5 and 10 cm dust layer models. For thicker dust mantles this relative deviation increases strongly and for dust mantle thicknesses larger than about 35 cm, there is no specie which is produced somewhere on the orbit with a relative deviation from the primitive composition smaller than 50\%. A thick dust layer is clearly an obstacle to the determination of the primitive composition of the nucleaus.

\paragraph{Day/night variations of the outgassing}

Unlike the 'nominal' amorphous model, the `dust mantle' models present continuous outgassing of all the species (including H$_2$O), without any fluctuation with the diurnal illumination cycle of the rotating nucleus. The sublimation interface of H$_2$O being deeper in the nucleus ($\geq$ 5 cm) than the diurnal thermal skin depth (about 0.7 cm), its temperature is only slightly affected by the day/night variations of insulation at the surface of the nucleus.

\subsection{Influence of other physical properties (thermal inertia, abundance and distribution of species) on the relative outgassing abundance of species \label{physicalproperties}}

The water ice structure and the existence of a dust mantle at the surface of cometary nuclei are not the only parameters that result in changes in the outgassing profiles of volatile species. We analyze hereafter the influence of some physical properties on the amorphous model (presented in the Sec.~\ref{sec:structure} and now named the `nominal model') such as the thermal inertia of the solid porous matrix of the nucleus, and the abundance and distribution between the `condensed' and `trapped' states of volatile molecules within the nucleus. All results are compared to the `nominal' model.

\begin{figure}
\begin{center}
\includegraphics[width=10.cm]{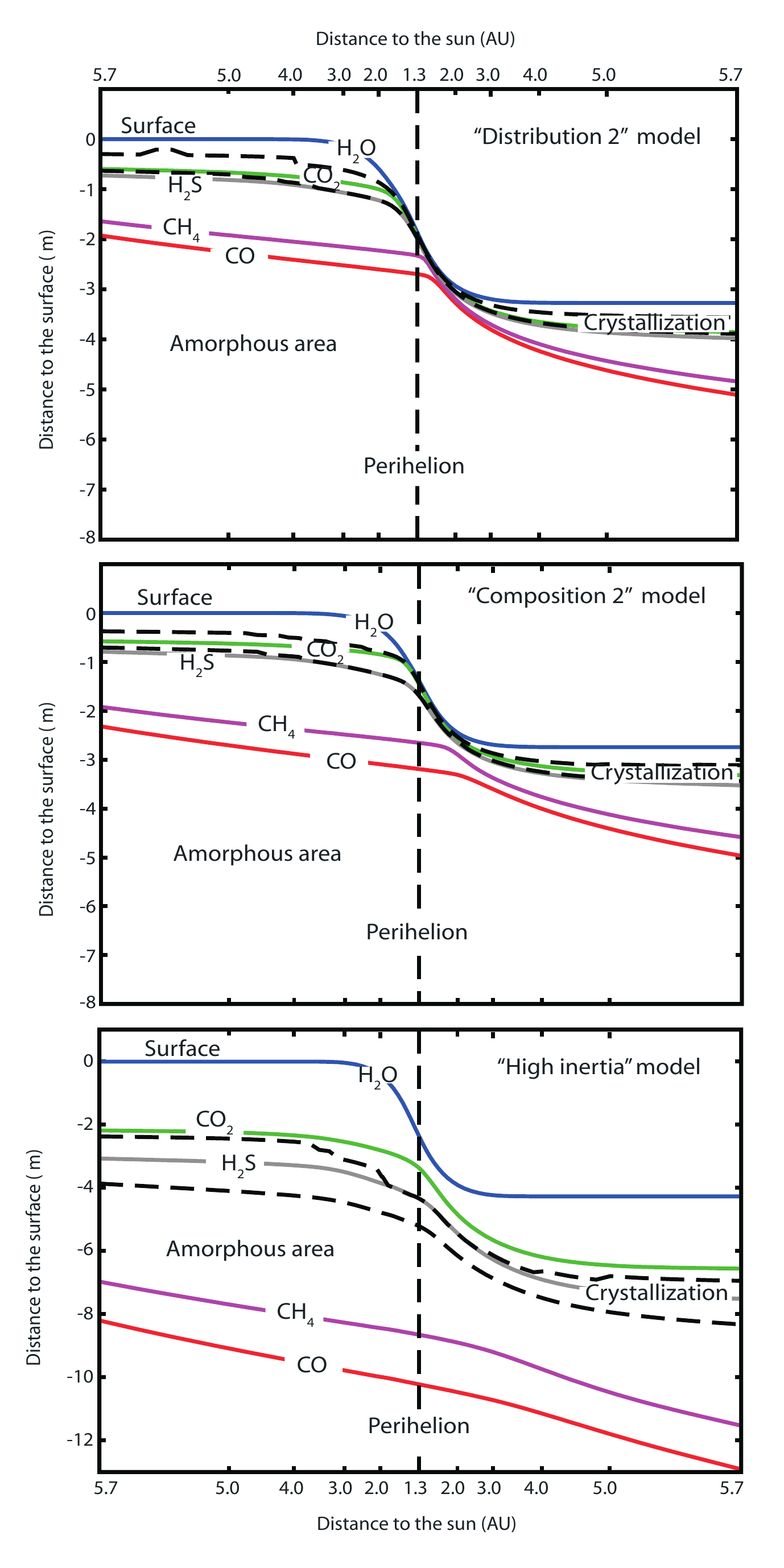}
\caption{\textbf{Physical properties models} - Physical differentiation of the comet nucleus for the amorphous models `distribution 2', `chemical composition 2', and `high inertia' as a function of the distance to the sun, during one revolution (the horizontal axis is linear in time). The vertical dashed line corresponds to perihelion. Calculations have been performed at latitude $\theta$ = 10$^\circ$. The thermal inertia is $\approx$ 90 W m$^{-2}$ K$^{-1}$ s$^{\frac{1}{2}}$ for the `high inertia' model and 30 W m$^{-2}$ K$^{-1}$ s$^{\frac{1}{2}}$ for the other models.}
\label{fig31}
\end{center}
\end{figure}

\paragraph{Physical differentiation of the nucleus}
Figure \ref{fig31} shows the physical differentiation of the nucleus as a function of the distance to the sun, during one revolution, for all the amorphous models (`nominal', `distribution 2', `chemical composition 2', and `high inertia'). The thermal inertia is $\approx$ 90 W m$^{-2}$ K$^{-1}$ s$^{1/2}$ for the `high inertia' model and 30 W m$^{-2}$ K$^{-1}$ s$^{\frac{1}{2}}$ for other models. For comparison, remember that the physical differentiation of the `nominal' model is shown in Fig.\ref{fig11}.

The `distribution 2' and chemical `composition 2' models show little differences of the physical differentiation of the nucleus compared to the `nominal' model. Mostly the chemical `composition 2' model presents slightly deeper interfaces of CO and CH$_4$ compared to the crystallization area. \textbf{This is due to lower abundances of these highly volatile species in the condensed state compared to the 'nominal' model.}
For the `high inertia' model, the interfaces of sublimation of all the species, as well as the crystallization interface, are situated deeper in the nucleus compared to the `nominal' model (see Fig.~\ref{fig11}, upper panel). The higher thermal inertia induces a physical differentiation over a depth of 13 m compared to about 5 m for the `nominal' model. At each perihelion passage, the average ablation of the surface reaches about 4 m in comparison to 3 m for the `nominal' model. \textbf{This higher differentiation of the nucleus is due to a higher thermal conductivity of the porous matrix which allows the thermal wave to penetrate deeper in the nucleus and thus push the sublimation and crystallization fronts to deeper layers.}

\begin{figure}
\begin{center}
\includegraphics[width=10.cm]{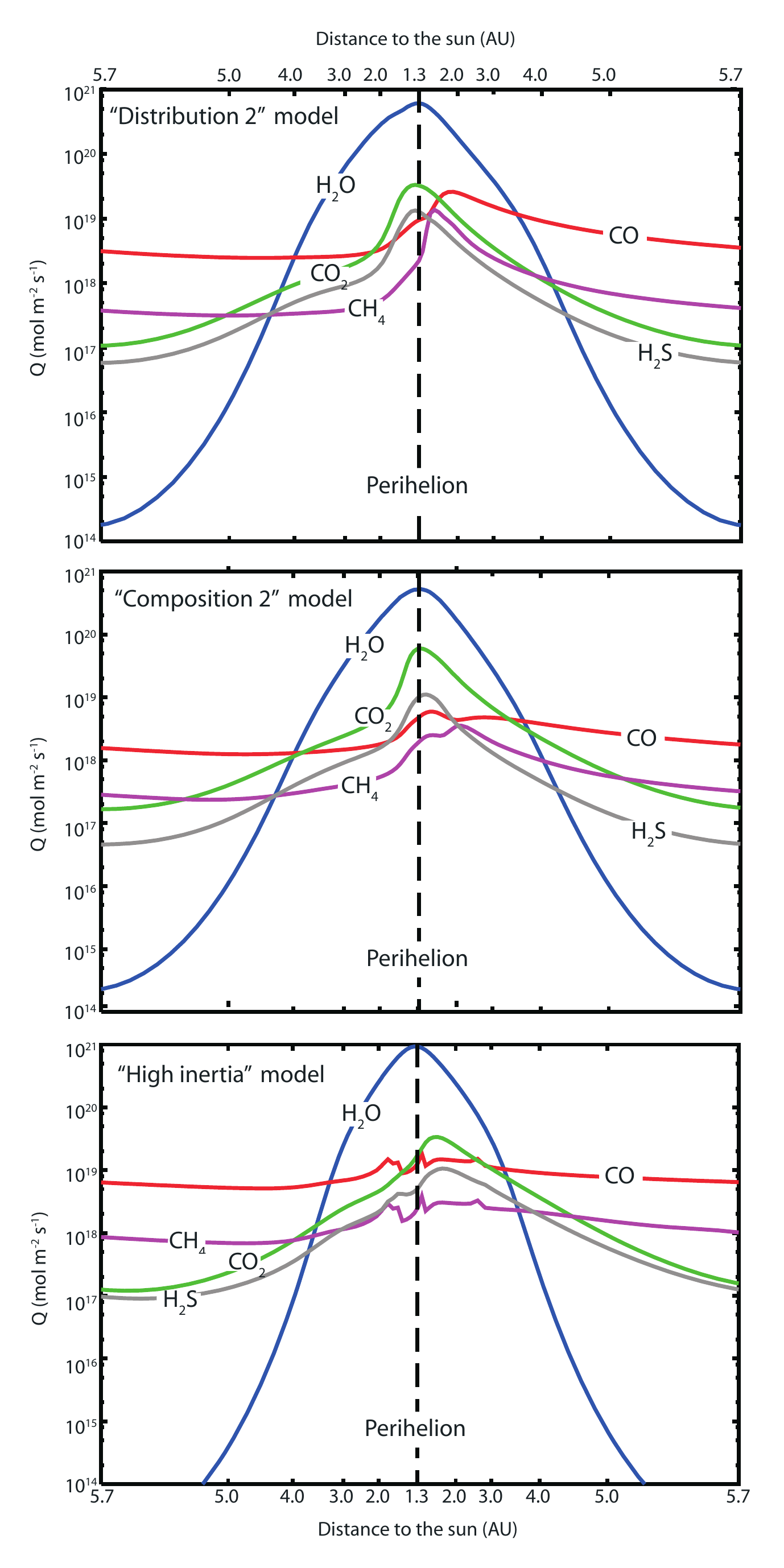}
\caption{\textbf{Physical properties models} - Comparison of the outgassing profiles, in molecules per second and per unit of surface area, of volatile molecules H$_2$O, CO, CO$_2$, CH$_4$, and H$_2$S for the different amorphous models as a function of the distance to the sun, during one revolution (the horizontal axis is linear in time). The vertical dashed line corresponds to perihelion.}
\label{fig32}
\end{center}
\end{figure}

\paragraph{Outgassing behavior}
Figure \ref{fig32} presents the outgassing profiles of H$_2$O, CO, CO$_2$, CH$_4$ and H$_2$S, in molecules per second and per unit of surface area, of the comet nucleus as a function of the distance to the sun, during one revolution, for the `distribution 2', chemical `composition 2', and `high inertia' models (all initially with fully amorphous ice). For comparison, remember that the outgassing profiles of species for the `nominal' model are shown in Fig.\ref{fig12}.

\begin{figure*}
\begin{center}
\includegraphics[width=7.cm, angle=-90]{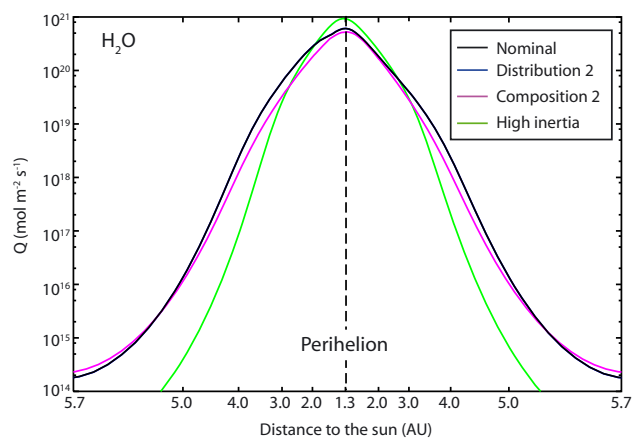}
\caption{\textbf{Physical properties models} - Comparison of the outgassing profiles, in molecules per second and per unit of surface, of H$_2$O for the amorphous models as a function of the distance to the sun, during one revolution (the axis is linear in time). The vertical dashed line corresponds to perihelion.}
\label{fig33}
\end{center}
\end{figure*}

\begin{figure*}
\begin{center}
\includegraphics[width=15.cm]{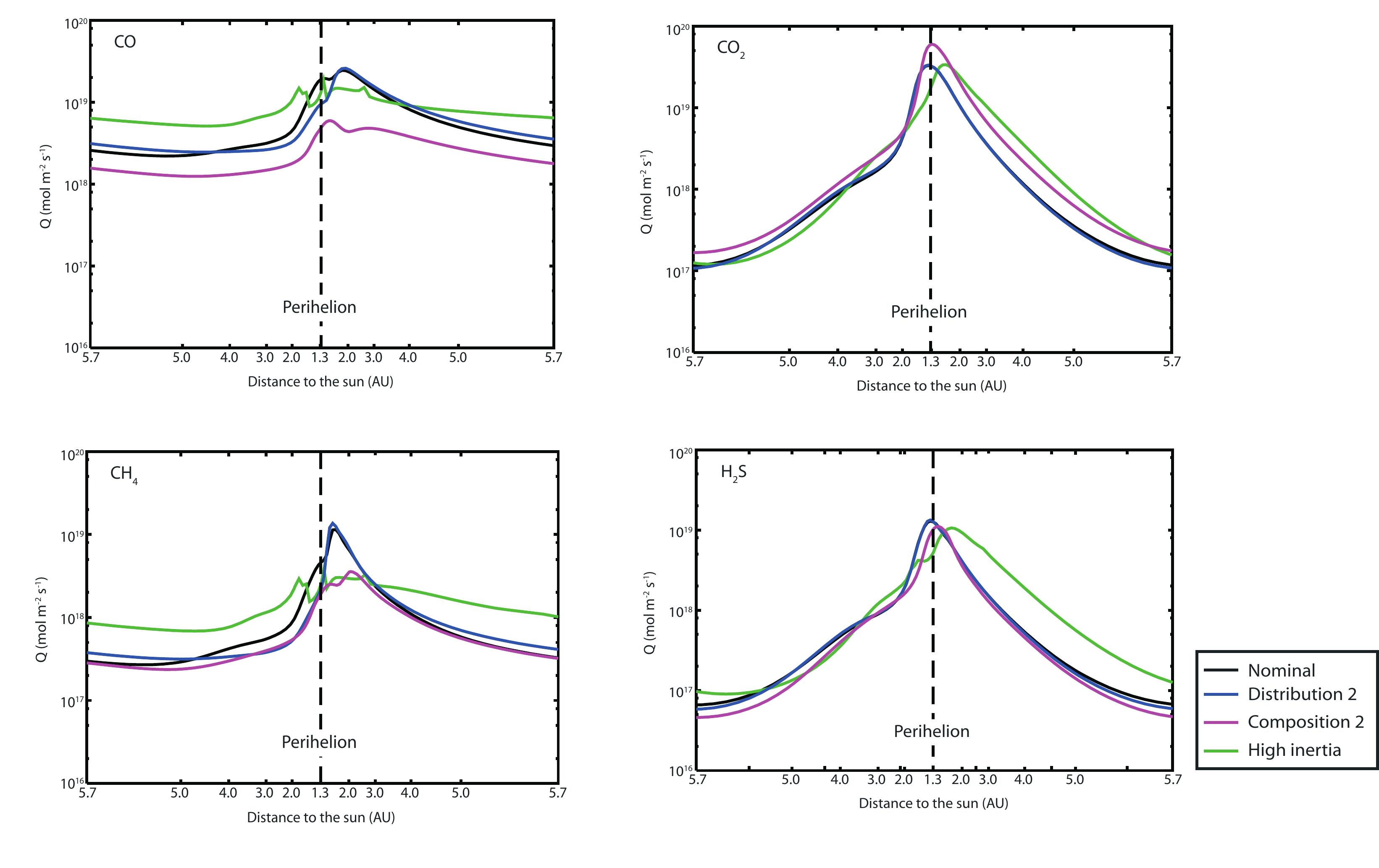}
\caption{\textbf{Physical properties models} - Same as in Fig.~\ref{fig33} but for CO, CO$_2$, CH$_4$, and H$_2$S.}
\label{fig34}
\end{center}
\end{figure*}

Figures \ref{fig33} and \ref{fig34} present the comparison of outgassing profiles, in molecules per second and per unit of surface area, of H$_2$O, and CO, CO$_2$, CH$_4$ and H$_2$S respectively, as a function of the distance to the sun, during one revolution for all the models (`distribution 2', chemical `composition 2', `high inertia', and 'nominal' models). 
The outgassing profiles of H$_2$O are mostly symmetric relative to perihelion although slightly shifted post-perihelion ($\approx$ 0.1 AU) in most models except the one with high inertia. 
For the high inertia model the profile of water production is steeper and with a stronger perihelion production, while a change in composition only very slightly affects the water sublimation over a wide distance around perihelion. On the other hand the distribution of species between the `trapped' and `condensed' states has no effect on the surface temperature of the icy nucleus and thus on its water production. 


With the `high inertia' model, the maximum production rates of the outgassing profiles of CO$_2$ and H$_2$S are slightly shifted post-perihelion ($\approx$ 0.13-0.2 AU) after the peak of H$_2$O production, and their wide pre-perihelion shoulder of production (around 4-3 AU) becomes more intense (see Fig.~\ref{fig34}). 
In Figure \ref{fig31} we can see that the stage of fast decrease of both the interfaces of these two species and of the crystallization front is shifted in time relative to the one of H$_2$O (centered on perihelion) leading to the observed shifts in their outgassing profiles. These sublimation and crystallization shifts comes from the much larger depths of their interfaces at aphelion (2-4 m compared to 20-70 cm in the low thermal inertia `nominal' model). The wide pre-perihelion production shoulder of CO$_2$ and H$_2$S are generated by the crystallization events starting slightly before 4-3 AU and releasing a large fraction of the primitive content of these molecules (60\% of which is trapped in amorphous ice) as well as heat that sublimates pure ices condensed around the cristallization front. 

For CO and CH$_4$, the post-perihelion increase of production observed for the `nominal' model (thermal inertia $\approx$ 30 W m$^{-2}$ K$^{-1}$ s$^{1/2}$) disappears in favor of smaller erratic fluctuations around perihelion, due to sporadic releases of these molecules ($\approx$ 30-40\% of which is trapped in amorphous ice) by a series of crystallization events. \textbf{For high thermal inertia, the crystallization of the amorphous layers occurs with energy provided from the surface (the sun) and from the crystallization itself as for the 'nominal' model, but with a lower fraction from the surface.}
CO$_2$ did not display such a series of small production peaks because crystallization occurs below the CO$_2$ ice interface (see Fig.~\ref{fig31}) and most CO$_2$ released during these events rapidly recondenses in the porous network of the nucleus. H$_2$S displays some of these peaks, but attenuated because crystallisation occurs just above its pure ice interface and part of the gas expelled by the amorphous ice should recondense as pure ice just below while another part diffuses towards the surface. 

For all the species (except H$_2$O), far away from the sun, the production rates are about a factor 1.5 to 3 higher than in the `nominal' model, except for CO$_2$ just after aphelion. Near the sun, they are mostly less than those for the `nominal' model. Due to the higher conductivity of the nucleus surface, the thermal energy penetrates deeper in to the nucleus and keeps its subsurface warmer over much of the short period orbit (\~ 6.6 years), while moderating the surface heating near perihelion.

So, with the same chemical composition, a change of the thermal inertia of the cometary nucleus, i.e. a change in the heat diffusion within the subsurface layers of the nucleus, induces significant changes in the differentiation of the nucleus, and consequently in the rates and outgassing profiles of all the volatile species.

The `distribution 2' model simulates a more efficient trapping of the less volatile CO$_2$ and H$_2$S molecules in amorphous ice (75-80\% of the total initial amount of these species, instead of 60\% in the `nominal' model) during the formation of the cometary material, at the expense of the trapping of the more volatile species CO and CH$_4$ (only 10\% of their total initial amount, instead of $\approx$ 40\%). The total amount of trapped species is also reduced from 8\% to 6.7\% in this model. 

This model presents outgassing profiles of CO$_2$ and H$_2$S indiscernible from the `nominal' model. As both the trapped and condensed fractions of these gases come from very near the surface (See Figures \ref{fig11} and \ref{fig31}), it is the surface ablation that mostly controls their production.
For the outgassing profiles of highly volatile species CO and CH$_4$, mostly the pre-perihelion outgassing and the first production peak (or shoulder) centered on perihelion are affected (Fig.~\ref{fig34}). They are strongly reduced compared to `nominal' model while the post-perihelion peak is almost unaffected. This pre-perihelion reduction is directly linked with the reduced amount of these volatile gases released by the crystallization of amorphous ice close to the surface, while the post-perihelion peak is mostly controlled by the surface ablation, both the crystallization and pure CO and CH$_4$ ices interfaces being now very close to the surface (Fig.~\ref{fig31}). There is also a noticeable increase in the production rates of these species far from the sun. 

The `composition 2' model simulates a comet nucleus strongly enriched in CO$_2$ (10\%) relative to CO (5\%) with a CO$_2$/CO ratio 4 times larger than the one of the `nominal' model (see Table \ref{param_ices2}). CH$_4$ has also a reduced abundance while H$_2$S stays at its nominal value. These changes in abundance are also reflected on the distribution between the condensed and trapped states with a larger amount of trapped CO$_2$ and H$_2$S to the detriment of CO and CH$_4$.
In this model, the shape of the outgassing profiles of CO$_2$ and H$_2$S are very similar compared to the `nominal' model. The absolute values of their production rates changes proportionally to the total abundance of the species in the nucleus. The only small changes noticed in their production shape can be attributed to changes in the distribution between the `condensed' and `trapped' states, as in the `distribution 2' model.
For CO and CH$_4$ the changes in the absolute values of their production rates are also directly linked with the changes of their total abundance in the nucleus. But the main effect is a significant shift ($\approx$ 0.2 AU for CH$_4$ and $\approx$ 0.8 AU for CO) of both their perihelion and post-perihelion production peaks. This shift is induced by a slowing down of the amorphous ice crystallization and of the surface ablation (towards the deep CO and CH$_4$ interfaces) due to the very high sublimation latent heat of pure CO$_2$ ice.

Finally, this set of results shows that extended observations (wide range of heliocentric distances) of the outgassing profiles of a series of species with different volatilities can help to constrain the thermal inertia of the nucleus as well as the abundance and distribution of species between the `trapped' and `condensed' states in an initially amorphous nucleus.

\begin{figure*}
\begin{center}
\includegraphics[width=15.cm]{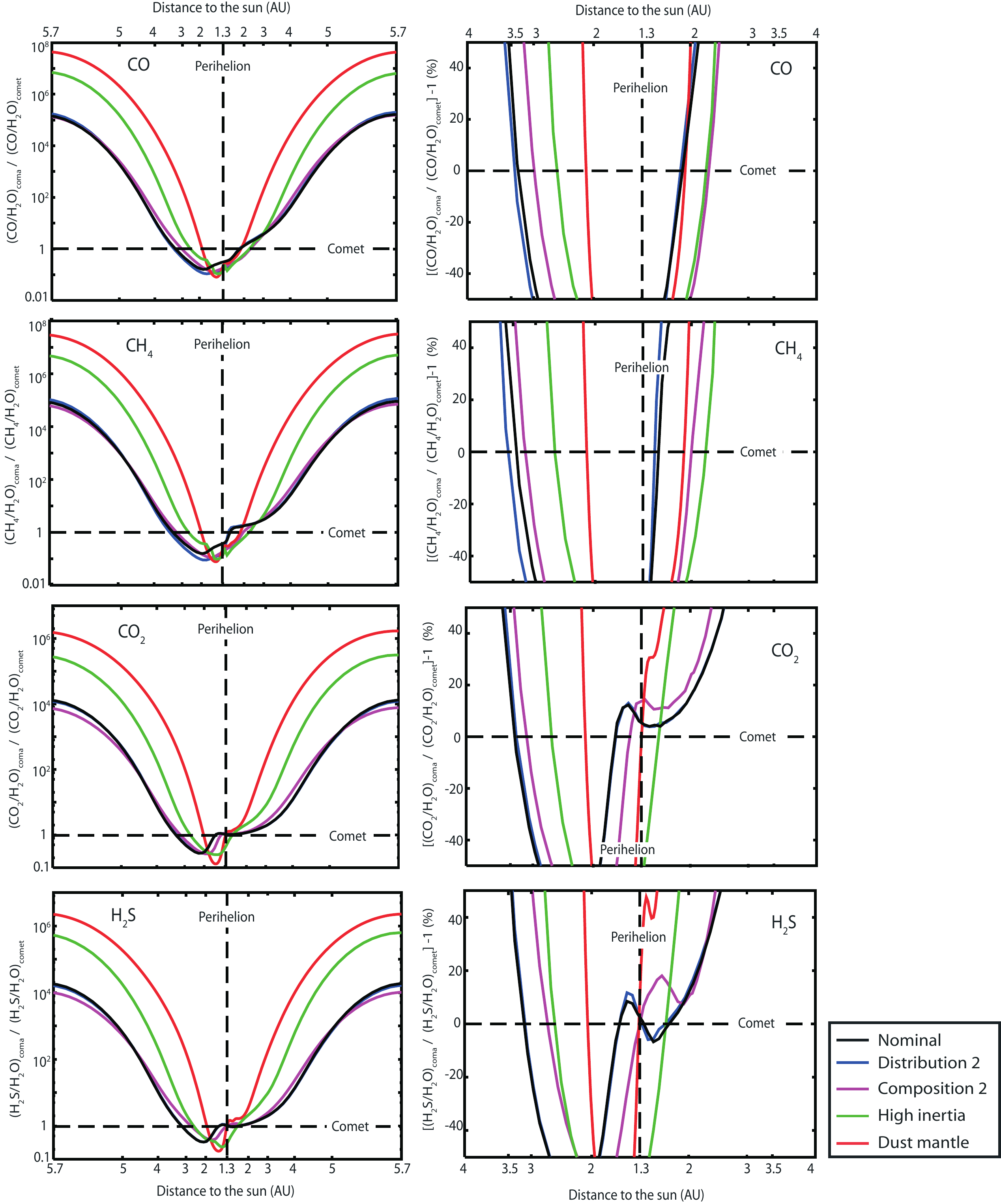}
\caption{\textbf{Physical properties models} - Ratio X/H$_2$O of the gas productions in the coma relative to the ratio in the primitive nucleus (left column) and deviation from the primitive composition of the nucleus (right column) for volatile species CO, CO$_2$, CH$_4$, and H$_2$S as a function of the distance to the sun. The value 1 (horizontal dashed line, left column) corresponds to the primitive abundance. The value '0' (horizontal dashed line, right column) corresponds to the case where no deviation is observed between the coma and the nucleus. The vertical dashed line corresponds to perihelion.}
\label{fig35}
\end{center}
\end{figure*}

\paragraph{Abundance of species in coma}
Figure \ref{fig35} presents the ratio $\frac{X}{H_2O}$ of the gas productions in the coma relative to the initial ratio $\frac{X}{H_2O}$ in the solid phase (sum of all solid phases) of the nucleus (left part of the figure) and the relative deviation of production of species from their initial nucleus abundances (right part) for species CO, CO$_2$, CH$_4$, and H$_2$S as a function of the distance to the sun, for the `nominal', `distribution 2', `composition 2', and 'high inertia' models. The value `1' (horizontal dashed line, left column of the figure) corresponds to the normalized initial nucleus abundance for each species. The value '0' (horizontal dashed line, right column of the figure) corresponds to the case where no deviation is observed between the coma and the nucleus. The vertical dashed line corresponds to perihelion. 
For comparison, outgassing profiles of species of 'nominal' (black lines) and 'dust mantle (5 cm of thickness, red lines)' models are also shown in the figure. 
As for the different ice phase models (see Fig.~\ref{fig14}), the relative abundances of all the species show variations of several orders of magnitude from aphelion to perihelion. For all the amorphous models, the less volatile molecules CO$_2$ and H$_2$S present relative abundances varying between $\approx$ 0.5 and 1 (0.1 and 1 for the `dust mantle' model) near perihelion while the more volatile species CO and CH$_4$ vary between $\approx$ 0.1 and 1. Far away from the sun, the relative abundances of species strongly depend on the model parameters, by several orders of magnitude. Most of these variations are due to the very strong sensitivity of the H$_2$O outgassing profile to thermal inertia, or to a lesser extent to composition (see Fig.~\ref{fig33}).
The relative outgassing of all the species are generally not symmetric relative to perihelion. 

Let us consider now the two places on the orbit where we previously noted a limited variation or the X/H$_2$O ratios for all species independently of the ice structure.
In the first place we can see that, around 3.5 AU before perihelion, the relative deviations of production of all species from their initial nucleus abundances are limited only when the thermal inertia is not changed. However for the `high inertia' model, the productions of all species increase relative to H$_2$O by one to six orders of magnitude, mostly due to a decrease of H$_2$O (see Fig.\ref{fig33}) by these amounts (decrease of the surface temperature).
On the other hand, at the second place on the orbit, from about 3 (pre-perihelion) to 2.5 AU (post-perihelion), the `high inertia' model displays X/H$_2$O ratios variations similar to the other models (less than 20\%).
This heliocentric distance post-perihelion (1.3-2.5 AU) seems the most interesting to derive approximate primordial relative abundances of cometary species, 'independently' of the internal differentiation processes of the nucleus, because of the lowest sensitivity of gas productions to thermal inertia (from ice or from a dust layer) and to the water ice structure at this particular place.

\paragraph{Day/night variations of the outgassing}

All amorphous models present the same behavior as the `nominal' model for the diurnal fluctuations of the gas productions of all species: only H$_2$O presents high variations with the diurnal illumination cycle of the rotating nucleus. The sublimation interface and area of crystallization, quite close to the surface, at about 8-10 cm depth, remain below several diurnal thermal skin depths (of about 1 cm) for `nominal', `distribution 2' and `composition 2' models. For the `high inertia' model, the diurnal thermal skin depth increases to about 3 cm but the sublimation interfaces of all the species are deeper. So, the sublimation interfaces of all the species in this model are poorly affected by diurnal variations. 

\section{Comparison to observations \label{observations}}

\begin{figure}
\begin{center}
\includegraphics[width=15.cm]{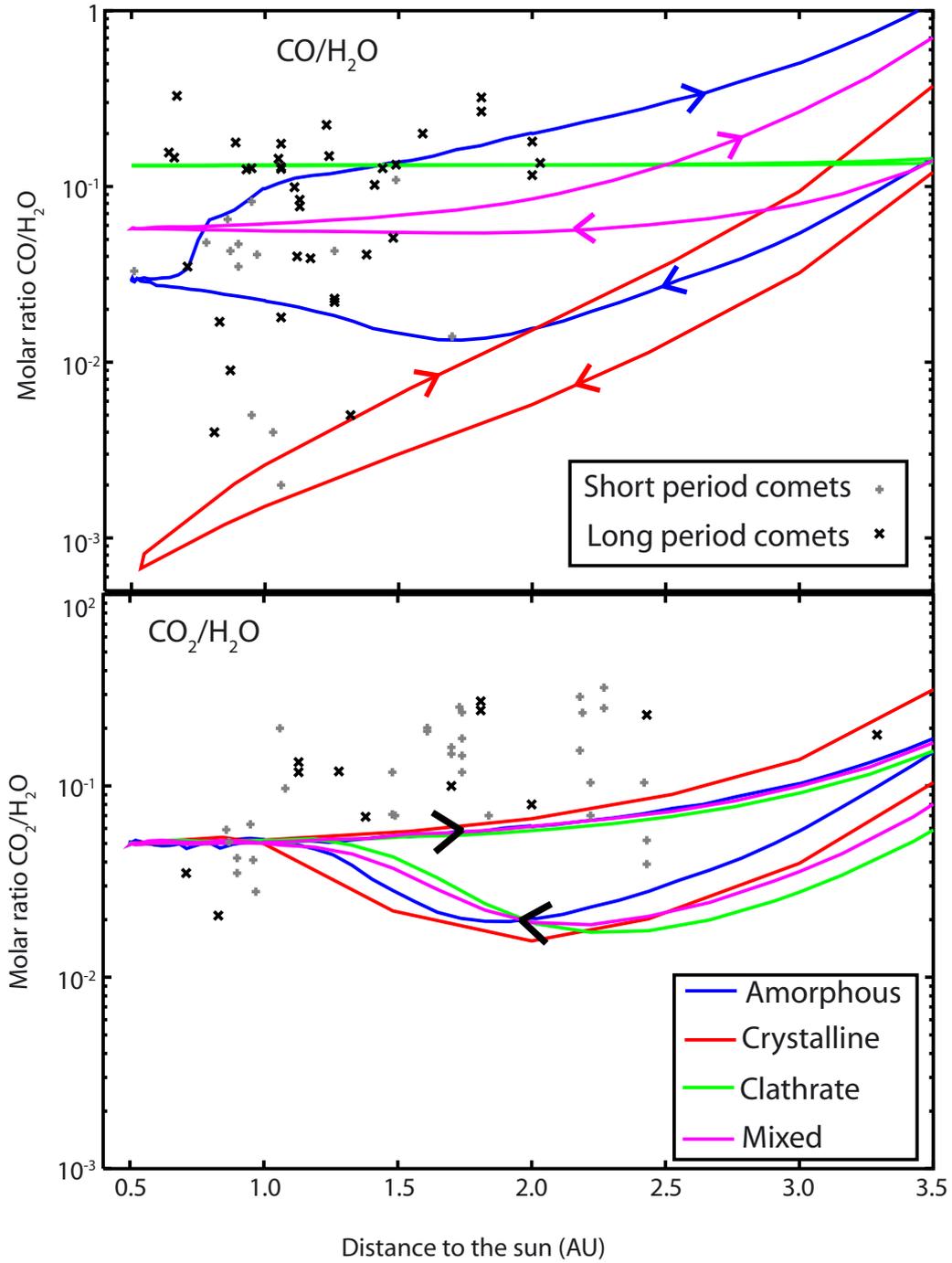}
\caption{Ratios CO/H$_2$O and CO$_2$/H$_2$O of the gas productions in the coma of comets as a function of the distance to the sun. Lines are simulations for comet 153P/Ikeya-Zhang for all `ice structure' models (amorphous, crystalline, clathrate, and mixed). Arrows represent the direction taken by the comet during its travel around the sun. Crosses correspond to observational data of short (gray +) and long-period (black x) comets (A'Hearn et al. 2012).}
\label{fig41}
\end{center}
\end{figure}

\begin{figure}
\begin{center}
\includegraphics[width=15.cm]{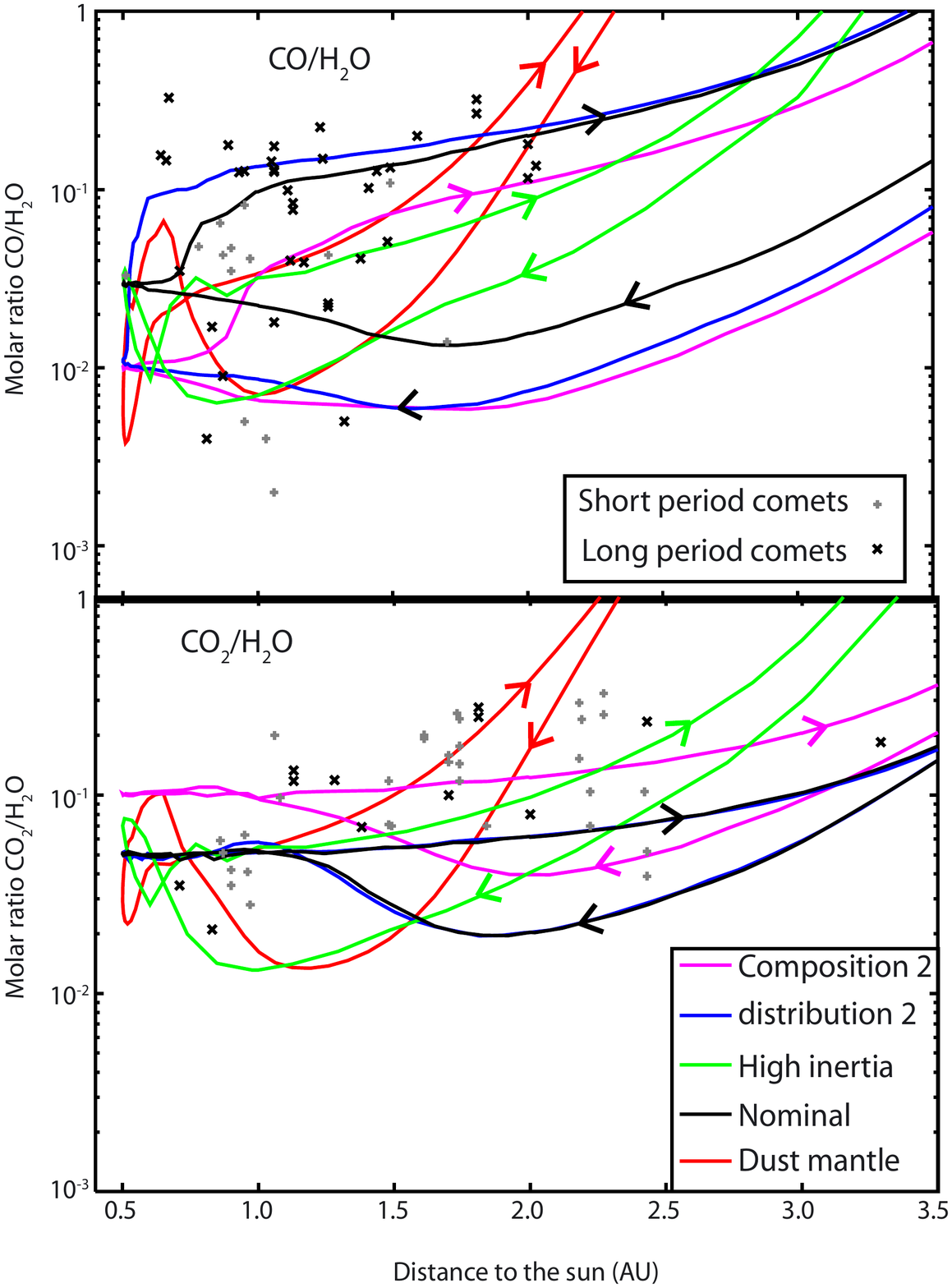}
\caption{Same as Fig.~\ref{fig41} but with all amorphous models (nominal, distribution 2, composition 2, high inertia, and dust mantle 5 cm thick).}
\label{fig42}
\end{center}
\end{figure}

In this section, we compare the results of the models to some observations of outgassing of comets. In order to fully cover the area of observational data of long and short period comets, we have chosen to take the short period comet 153P/Ikeya-Zhang (period of 366 years) that has a perihelion at 0.5 AU (the smallest heliocentric distance of the observational data from A'Hearn et al. 2012). In the models we run for 153P/Ikeya-Zhang the deviation of the gas production ratio X/H$_2$O in the coma relative to the initial ratio X/H$_2$O in the nucleus does not differ significantly compared to the ones of 67P/C-G comet. The orbital parameters of this comet are provided in Tab.~\ref{paramss}. Its radius (2 km) has been arbitrarily chosen to be similar to those of JFCs hovering around 2-3 km (see Weiler et al. 2011; Snodgrass et al. 2011). Note that results are shown during one revolution and
after 5000 years of revolution around the sun (about 14 revolutions). \textbf{Note also that we ran the models during 29 orbits (10000 years) and that results don't change significantly during this time scale.}
The thermodynamical parameters used for the following study are the same as the one for the nucleus 67P/C-G.
Figure \ref{fig41} and \ref{fig42} present the ratios $\frac{CO}{H_2O}$ and $\frac{CO_2}{H_2O}$ as a function of the distance to the sun, from perihelion to 3.5 AU, for all models (amorphous, crystalline, clathrate, mixed, distribution 2, composition 2, high inertia, and dust mantle 5 cm thick). Crosses represent the observational data of short (gray) and long (black) period comets from A'Hearn et al. (2012). Arrows represent the direction taken by the comet during its travel around the sun.

\begin{figure}
\begin{center}
\includegraphics[width=15.cm]{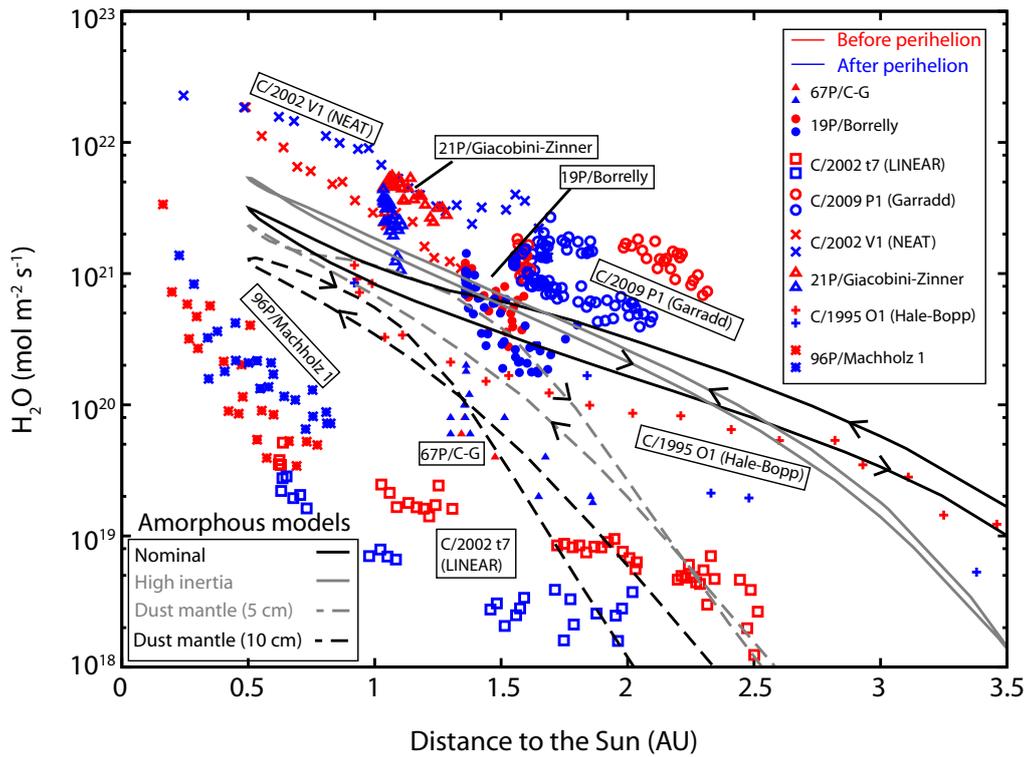}
\caption{Molar production $Q(r)$ (mol m$^{-2}$s$^{-1}$) of H$_2$O per unit of cometary surface area and per second as a function of the distance to the sun. \textbf{Red} (before perihelion passage) and \textbf{blue} (after perihelion passage) points represent the observational data of comets C/1995 O1 (Hale-Bopp), C/2002 V1 (NEAT), C/2002 T7 (LINEAR), C/2009 P1 (Garradd), 19P/Borrelly, 21P/Giacobini-Zinner, 67PC-G, and 96P/Machholz 1 for which nucleus sizes and H$_2$O production data exist. Lines represent the H$_2$O production rate for comet 153P/Ikeya-Zhang from the amorphous 'nominal', 'high inertia' and 'dust mantle' (5 cm and 10 cm of thickness) models. Arrows represent the direction taken by the comet during its travel around the sun.}
\label{fig43}
\end{center}
\end{figure}

Observational data of the ratios CO/H$_2$O and CO$_2$/H$_2$O show high dispersions from 2 10$^{-3}$ to 0.33 and from 2 10$^{-2}$ to 0.33 respectively. As for the 67P/C-G comet (see Fig.~\ref{fig14}), the numerical simulations for comet 153P/Ikeya-Zhang show large variations of the production profiles and relative abundances of the highly volatile species CO as a function of the structure of water ice (see Fig.~\ref{fig41}). All models, excepted the 'clathrate' model, present asymmetric outgassing behavior of the relative abundance CO/H$_2$O about the perihelion passage from 0.5 to $\pm$3.5 AU. 
For all the different types of ice structure models, the relative abundance CO/H$_2$O is higher after perihelion than before (the reverse is true for the outgassing of H$_2$O, see Fig.~\ref{fig43}). The crystalline structure presents very low values ($<$ 1\%) of the relative abundance CO/H$_2$O in the coma, especially at solar distances less than 2 AU. A much higher abundance of CO (one order of magnitude) in the nucleus of the comets than in the model (10\%) would be required to reproduce most of the observations. Amorphous and clathrate models show higher values, more consistent with the majority of the observations. However, the amorphous model presents a strongly assymetric outgassing behavior about the perihelion passage while the clathrate nucleus has a continuous and highly symmetric (about perihelion) relative abundance of CO/H$_2$O. 

This comparison with the observational data, both for long and short period comets suggests that comets could be mainly composed of these 2 last structures of water ice: amorphous ice and clathrate. And that crystalline ice is mostly excluded, except maybe in a few comets producing less than a few percents CO or for comets with larger CO abundance in the nucleus.
Increasing or decreasing the primitive CO abundance in the nucleus for all models by no more than a factor of 2 (i.e. covering the 5-20\% range) allows one to fully cover the range of observations, except the observation of 33\% CO in comet C/1996 B2 Yakutake at 0.67 AU. By changing the physical properties of the amorphous ('nominal') model (see Fig.~\ref{fig42}), the relative abundance CO/H$_2$O of all amorphous models cover almost all the area of observational data from 0.5 to 2 AU. As for the 67P/Churyumov-Gerasimenko comet (see Fig.~\ref{fig14}), the thermal inertia, the distribution of species between the `trapped' and `condensed' states, the initial abundance of species, and the physical differentiation (dust mantle) of the nucleus change the production profile and relative abundance of CO/H$_2$O in the coma of comets, allowing to span almost the whole range of observational data.

The main argument in favor of the amorphous ice assumption in comets should be the asymmetry and the shift of the outgassing of highly volatile molecules: comets can present strong asymmetries of gas production with strong chemical inhomogeneities (Lederer \& Campins 2002; Bockel\'ee-Morvan et al. 2009). The strong asymmetry of the relative abundance CO/H$_2$O in the coma of amorphous models comes from both the asymmetry of H$_2$O and CO outgassing. Remember that the outgassing profile of CO is similar at perihelion to the one of H$_2$O for clathrate-rich models (see Fig.~\ref{fig12}), leading to a constant CO/H$_2$O ratio near perihelion (see Fig.~\ref{fig41}). Unfortunately, the data on the outgassing of species over a significant range of heliocentric distances are very scarce (very limited number of comets) and concern mainly H$_2$O. 
Figure~\ref{fig43} shows the H$_2$O molar production curves (mol m$^{-2}$s$^{-1}$), per unit of cometary surface area and per second, of some \textbf{short and} long-period comets compared to amorphous models. \textbf{Red (before perihelion passage) and blue (after perihelion passage)} dots represent the observational data of comets C/1995 O1 (Hale-Bopp), C/2002 V1 (NEAT), C/2002 T7 (LINEAR), C/2009 P1 (Garradd), 19P/Borrelly, 21P/Giacobini-Zinner, 67PC-G, and 96P/Machholz 1, for which estimations of nucleus sizes (Fern´andez 2002; Sosa \& Fern´andez 2011; Weiler et al. 2011; Combi et al. 2013) and H$_2$O productions (Biver et al. 2002; Combi et al. 2009, 2011a, 2013) exist over a significant range of heliocentric distances. Lines represent the H$_2$O production from the amorphous 'nominal', 'high inertia' and 'dust mantle' (5 cm and 10 cm thick) models. Arrows represent the direction taken by the comet during its travel around the sun. 
\textbf{Note that comets without dust mantle can have a higher rate of H$_2$O production 'before' perihelion passage compared to 'after'. As shown in Fig.~\ref{fig11}, the ablation of the surface removes 3 meters of ices, and reaches the interfaces of sublimation of the less volatile species CO$_2$ and H$_2$S which take also energy for their sublimation. 
 Although the thermal wave penetrates several tens of centimeters in subsurface layers during one orbit, the ablation of the surface removes these layers and thus the energy stored in them during the travel of the comet: the comet becomes then as thermally new, with cold subsurface layers. It results that the gradient of temperature inside the nucleus becomes higher 'after' than 'before' perihelion: the energy transferred from the surface to subsurface layers by thermal conductivity increases (see Eq.~\ref{boundary_heat_surf}). For a same heliocentric distance, this results in a decrease of the surface temperature (and hence the pressure of sublimation of H$_2$O) 'after' perihelion passage compared to 'before'. Note that this behavior is opposite when the comet is covered by a permanent dust mantle. In such a case, the thermal wave penetrates the subsurface layers without a big ablation of the surface during the orbit since the sublimation front of H$_2$O ice is deeper than in models without dust mantle. The water production rate becomes thus higher 'after' than 'before' perihelion passage. This suggests that older comets, with a large fraction of their surface covered by a dust mantle, are more active post-perihelion than pre-perihelion.}

\textbf{The H$_2$O molar productions from comet nuclei are sometimes higher, sometimes lower than the theoretical data from our set of models: their outgassing data} show a dispersion of several orders of magnitude from one comet to another. These discrepancies can be explained \textbf{by different physico-chemical conditions in nuclei such as the ice to dust mass ratio $J_{dust}$ and/or the porosity $\Psi$ which change both the thermal inertia, and the water fraction, and thus the water production rate. The change of obliquity and rotational period between comets could also explain some of these discrepancies: a lower/higher rotational period $P_r$ of the nucleus induces a lower/higher rate of H$_2$O production.} \textbf{Differences can also be explained by} different fractions of active surface area of nuclei (i.e., dust-free or with only a thin dust layer; the local dust layer being built by physical differentiation and can vary depending on the dynamical history of nuclei) \textbf{or by a thicker dust mantle}, by an initial depletion of volatile species in some comets due to different areas of formation in the early Solar System, and by the uncertainties about the physical characteristics (shape, size, ...) of the nuclei: their sizes and fractions of active surface area were mainly computed from their estimated absolute nucleus magnitudes and their water production rates (Lamy et al. 2004; Tancredi et al. 2006; Snodgrass et al. 2011). Thus, the estimations of the size of nuclei can vary by factors of 2-3, such as for Hale-Bopp with a radius varying from 20 to 50 km following the methods used (Weaver \& Lamy 1997, Fernandez et al. 2002, McCarthy et al. 2007, Sosa \& Fern´andez 2011).
In addition, the high values of H$_2$O production for the long-period comets C/2002 V1 (NEAT) and C/2009 P1 (Garradd) could be also due to the sublimation of icy materials not only from the surface of the nucleus, but also from grains ejected in the coma, producing an extended source of gas and generating an activity larger than expected for a 100\% active surface nucleus (Combi et al. 2011b; Sosa \& Fern\'andez 2011).

Our amorphous models clearly show, whatever the model, the asymmetry of the H$_2$O production relative to perihelion. For all the models, the H$_2$O production is higher before perihelion than after (the reverse is true for the relative abundance CO/H$_2$O, see Fig.~\ref{fig41}). Note that the asymmetry of the H$_2$O outgassing is reversed for the 'dust mantle' model between 1 and 1.5 AU: the H$_2$O production becomes higher after perihelion than before (shift of the production peak, see Fig.\ref{fig22}).
For most of the comets, the H$_2$O production is higher before perihelion than after, excepted the comets 67P/C-G, 96P/Machholz 1, and C/2002 V1 (NEAT) which have reversed outgassing behaviors: the H$_2$O production is higher after perihelion than before. For comet C/2002 V1 (NEAT), Combi et al. (2011a) suggested a strong seasonal variation during orbit with regions near what was the mostly the dark pole before perihelion being exposed to strong sunlight for the first time at and after perihelion. 
For the observed comets, their water production rate varies rather irregularly during the approach of the nucleus to the sun (Combi et al. 2011b). This behavior can be explained by a strong compositional inhomogeneity of the surface and subsurface of the rotating nuclei (physical differentiation of nuclei), with some surface area partially covered with dust mantles and other with ices fully exposed to sunlight.

For the outgassing of the less volatile specie CO$_2$, all models (with 5\% CO$_2$ in the nucleus) produce rather low values of the relative abundance CO$_2$/H$_2$O in the coma compared to observations (see Fig.~\ref{fig41}). With this molecule, it is difficult to distinguish the different structures of water ice in comets since it is mainly the erosion of the surface of the nucleus (H$_2$O) that triggers the sublimation of this species around perihelion and that controls its outgassing rate. Remember that the relative CO$_2$/H$_2$O abundance in the coma is similar to the primitive composition of comets (deviation less than 25\%) from shortly before perihelion passage, 2 AU out to 2 AU post-perihelion. This means that the abundances of CO$_2$ in comets may be much higher than the one adopted in our study (5\%).   
By changing its abundance in the amorphous model (see Fig.~\ref{fig42}), or the thermal inertia or the physical differentiation of the nucleus (dust mantle), the production profile and the CO$_2$/H$_2$O ratio in the coma of comets change by 1 to 2 orders of magnitude, allowing one to cover almost the whole range of observational data. No more than a factor 2-3 increase of the primitive CO$_2$ abundance in the nucleus (i.e. over the 5-15\% range) should be necessary to easily model all CO$_2$/H$_2$O production ratios. It is interesting to note that the presence of a thin (5 cm) dust mantle produces an increase of the CO$_2$/H$_2$O ratio after 1.5 AU compared to the initial composition of the nucleus (Fig.~\ref{fig24}), as frequently observed (Fig.~\ref{fig42}).

This discussion is limited to the global variability of CO and CO$_2$ relative abundances in observations, and to the H$_2$O outgassing profile of some long-period comets but it is clear that an analysis comet by comet taking into account the production curves of the different molecules as a function of the distance to the sun is the only valid way to assess the primordial composition and structure of its nucleus. However such production curves over a significant range of heliocentric distances are very sparse and are currently mostly limited to the long-period Hale-Bopp comet for nine molecules and distances varying from 0.9 to 15 AU. Such an analysis is currently under way for Hale-Bopp for which we will discuss all the possible ice structures and chemical compositions in the nucleus which can reproduce the set of outgassing observational data.

\section{Conclusion \label{discussion}}

The chemical composition, water ice structure, thickness of dust mantle and thermal inertia of the icy nucleus of comets remain to be determined by future observations such as the Rosetta spacecraft mission in order to obtain information about the thermodynamic conditions of formation of the primitive solar system.

We have shown in this study that using a model of a cometary nucleus taking into account all phase changes of the icy materials to analyze the outgassing behavior of volatile species (production profile, amplitude, day/night variations, ...) in the coma of comets can help to constrain the physical properties of cometary nuclei, and hence the conditions of formation of these objects. The outgassing profiles and rate of productions of the volatile species allow one to constrain the water ice structure, the distribution of volatile molecules between the `condensed' and `trapped' states in the nucleus, as well as the thermal inertia of the nucleus.
In particular, for highly volatile molecules such as CO and CH$_4$, the number and position of peaks of gas production around perihelion, and the day/night variation of outgassing rates are strong indicators of the origin of the species in the nucleus (condensed, trapped deeply or near the surface in water ice or clathrate) and of the structure of water ice, major witnesses of the thermodynamic conditions of cometary ice formation in the protoplanetary disc. Interestingly, a nucleus containing clathrates presents an increased outgassing of all species, except CO$_2$ (because not trapped), on the day side of the nucleus while the other water ice structures show no diurnal outgassing cycle. A much lower thermal inertia ($\leq$ 10 W m$^{-2}$ K$^{-1}$ s$^{\frac{1}{2}}$) than the nominal value (30 W m$^{-2}$ K$^{-1}$ s$^{\frac{1}{2}}$) would be required to show some day/night variations of some species (those which are the closest to the surface) with the amorphous and crystalline models. In addition the presence of a dust crust has the effect to dampen or remove any fluctuation of all species (including H$_2$O).

We have shown that the abundance of volatile molecules (relative to H$_2$O) released from the interiors of nuclei vary by several orders of magnitude as a function of the distance to the sun, whatever are the physical properties (thermal inertia, physical differentiation, water ice structure, chemical composition, ...) of the nucleus. The volatility of the molecules, their abundance and distribution between the `trapped' and `condensed' states, the thermal inertia, the structure of water ice in the nucleus, and the presence of a permanent dust mantle at the surface of comets change the orbital profiles of relative abundance of species in the coma. The consequence is that observations of molecular abundances in the coma of these objects and conclusions about their chemical composition could be in error without taking into account models of cometary nuclei and assumptions on their physical properties (thermal inertia, water ice structure and gas trapping). Near perihelion, only comets mostly made of clathrate present relative abundances of species released in the coma that closely mimic the primitive composition of the nucleus. \textbf{However, for a nucleus containing partly emtpy clathrate structures (i.e. with a hydrate number higher than 6) the near perihelion outgassing rate of species initially trapped in clathrates should fall down near amorphous structure values.}

Cometary nuclei made of the other water ice structures, amorphous or crystalline ices, produce relative (to H$_2$O) abundances similar to the primitive composition of the nucleus only for the less volatile species CO$_2$ and H$_2$S and only around the perihelion passage (appriximately in the range -3-2 to +2-3 AU), and under the conditions that the thermal inertia of nuclei is low ($\leq$ 30 W m$^{-2}$ K$^{-1}$ s$^{1/2}$) and that they are not fully covered by a dust mantle.
Such comets release systematically lower relative abundances of the highly volatile CO and CH$_4$ molecules by up to one order of magnitude in the pre-perihelion part of their orbits (from ~3 AU to perihelion, see Figures \ref{fig14} and \ref{fig35}) except when there is a large dust layer ($\geq$ 30 cm) at the surface which strongly decreases the water production and consequently increases all the molecular X/H$_2$O ratios by several orders of magnitude (Fig.~\ref{fig24}).

However we noticed that around 3.5 AU pre-perihelion, all species are produced with a relative deviation of less than 50\% from their initial nucleus abundances, whatever is the short period comet model except when there is a dust mantle, for which the relative deviation increases strongly. This deviation is even much lower (less than 10\%) for low volatility molecules such as CO$_2$ and H$_2$S and occurs over a wider heliocentric range around perihelion, from -1.5 to +1.5 AU. These features could be used to help infer absolute abundances from the observation of gas released by active crust-free spots on the nucleus. 

Note that split comets such as the comets C/1999 S4 (LINEAR) (Bockel\'ee-Morvan et al. 2001) and 73P/Schwassmann-Wachmann 3 (Dello Russo et al.2007) or outbursts events could help to determine the primitive abundance of volatile species and structure of water ice.
Note also that this study was realized for chemically homogeneous comets, and tested with only one value of porosity, for spherical nucleus without obliquity, and only for 2 perihelion positions (0.5 and 1.2 AU). A non spherical nucleus composed of planetesimals with different chemical compositions (because formed in different areas of the protoplanetary disc, or that undergo chemical evolution/differentiation during their storage in cold areas far from the sun) should not present the general outgassing behavior presented in this study. However, locally, our model could fit the outgassing per unit of surface area and help to determine the local composition below the surface. In addition, by changing the perihelion position for higher value, we should see higher changes from one revolution to another, mainly because water sublimation should decrease significantly up to the fade-out of the erosion of the comet. In this last case, only highly volatile species should continue to sublimate and the predictions given in this study should be changed by substituting H$_2$O by a higher volatile species in order to predict relative abundances (not relative to H$_2$O) and structure of water ice in the nucleus. To determine the chemical composition and the structure of water ice in nuclei, future studies need to analyze the abundances of species (relative to H2O and to other species also), but also the shape of the outgassing profiles of volatile species, as well as the day/night outgassing that could help to distinguish a particular model. However, currently the answer is not trivial since a lot of parameters are free and several models could match data. 

The future measurements of some of the surface and subsurface properties of the 67P/Churyumov-Gerasimenko target nucleus, thanks to a set of remote and `in-situ' instruments (MIRO, VIRTIS, CONSERT, OSIRIS, MIDAS, RSI...) of the Rosetta spacecraft mission, should better constrain the values of some currently quasi-free physical parameters of the models and thus narrow the range of gas production profiles. In particular, knowledge of the subsurface thermal conductivity and porosity, i.e. the thermal inertia, of the nucleus, and of the thickness of the dust mantle at the nucleus surface will considerably narrow our parameter ranges. The future observations by the Rosetta mission instruments (VIRTIS, MIRO, ALICE, ROSINA, ...) of the outgassing profiles of various species released by the nucleus, and of their relative abundance (to H$_2$O), from $\approx$ 3 AU pre-perihelion to $\approx$ 3 AU post-perihelion should allow us to better constrain the ice structure and the abundances of the volatile species inside the nucleus, with the hope to be able to trace the primordial composition of the comet and finally the thermodynamic conditions of formation of the cometary material in the early solar system.

\paragraph{Acknowledgements}
The authors thank the detailed review reports by Jacques Crovisier and Paul Weissman that helped us to improve substantially the presentation of our results.
This work has been supported by the French 'Centre National d'Etudes Spatiales' (CNES), by the Swiss National Science Foundation and the Center for Space and Habitability of the University of Bern.
All the computations presented in this paper were performed at the Service Commun de Calcul Intensif de l'Observatoire de Grenoble (SCCI).

\end{document}